\documentclass[%
 reprint,
 superscriptaddress,
 amsmath,amssymb,
 aps,
 prb,
]{revtex4-2}

\usepackage{physics}
\usepackage{graphicx}
\usepackage{dcolumn}
\usepackage{bm}
\usepackage{booktabs} 
\usepackage{amssymb}

\usepackage{xcolor}
\usepackage{tikz}
\usetikzlibrary{calc,positioning,backgrounds,decorations.pathmorphing,arrows.meta,shapes.geometric,fadings,fit}

\definecolor{deeppurple}{HTML}{7b6ef0} 
\usepackage{hyperref}
\hypersetup{colorlinks=true,allcolors=deeppurple}

\begin{document}

\title{ILMlip: fast and universal machine-learning interatomic potentials}

\author{Jonas H{\"a}nseroth}
\email{jonas.haenseroth@tu-ilmenau.de}
\affiliation{Theoretical Solid State Physics, Institute of Physics, Technische Universit{\"a}t Ilmenau, 98693 Ilmenau, Germany}

\author{Christian Dre{\ss}ler}
\affiliation{Theoretical Solid State Physics, Institute of Physics, Technische Universit{\"a}t Ilmenau, 98693 Ilmenau, Germany}

\date{\today}

\begin{abstract}
Large-scale pretrained machine-learned interatomic potentials keep growing more accurate as new models appear at a rapid pace, and more expensive: the current leader of Matbench Discovery as of November 2026 (F1 score 0.935) has $9.9\times10^7$ parameters and took 3.6 thousand GPU-hours to train.
The ILMlip potentials presented here are optimised for speed as well as zero-shot accuracy.
They are equivariant, conservative message-passing potentials in which every component is arranged for throughput: little work per pair of atoms and as much as needed per atom.
Through ASE on one GPU, they run molecular dynamics up to about $10^3$~times faster than state-of-the-art message-passing potentials such as MACE (medium) and up to five times faster than NEP89 and DPA-4.1C.
At the same time, six sizes with $1.4\times10^5$ to $2.8\times10^7$ parameters, pretrained on materials, molecules and solid acids, reach F1 scores of 0.75 to 0.91 on Matbench Discovery, placing the largest within 0.03 of the current leader.
They therefore suit high-throughput, large or long simulations, and, being transferable, they are strong starting points for fine-tuning.
Fine-tuned to six materials, the model of size S comes close to the ab initio force errors of a fine-tuned MACE, so it reaches reference-level accuracy at about 50 ns per atom and step (Cu fcc).
Their components can be adopted separately, as a starting point for architectures with even better ratios of speed to accuracy.
The ILMlip implementation is available at \url{https://github.com/jhaens/ilmlip}.
\end{abstract}

\maketitle

\section{Introduction}
\label{sec:intro}

Machine-learned interatomic potentials (MLIPs) predict forces and energies close to the accuracy of the electronic-structure method they were trained on, at a cost that grows in principle linearly with the number of atoms~\cite{behler2007,bartok2010,zhang2018,drautz2019,batzner2022,mace}. 
Large-scale pretrained MLIPs, so-called universal or foundation MLIPs (UMLIPs), trained on databases that cover most of the periodic table~\cite{mp_1,mp_2,chgnet,omat24,alexandria,omol,matpes}, offer such a potential to users who have no training set of their own~\cite{m3gnet,chgnet,macemp0,sevennet,mattersim,orbv3,grace_2,esen,uma,pet_mad,nep89}. 
Their accuracy has risen quickly, and mostly through size. 
On \textsc{Matbench} Discovery~\cite{matbench}, where a model must decide which of about $250$~thousand hypothetical crystals are stable, the F1 score of this decision ($1$ is perfect) went from $0.57$ to $0.94$ within about three years. 
The current leader as of November 2026, \textsc{DPA-4.1-Max-OAM-PFT}~\cite{dpa4}, has $9.9\times10^7$~parameters and was trained for $3.6$~thousand GPU-hours (\textsc{Nvidia} RTX PRO 6000 Blackwell Server Edition); \textsc{PET-OAM-XL}~\cite{pet_oam}, another top-performing pretrained MLIP ($7.2\times10^8$~parameters), was trained for over $20$~thousand GPU-hours (\textsc{Nvidia} H100). 
Their size is paid for again in every step of every simulation that uses them.

Much of what is simulated does not need the top of this scale. 
A high-throughput screening study, a simulation cell large enough to hold an interface or a trajectory long enough to show a rare event is limited by the number of force calls it can afford, and there a model that is nearly as accurate and far cheaper reaches further than the most accurate one for a fixed computational budget. 
Where a property must be trusted, a universal model applied as it is, zero-shot, is rarely the end point; the canonical route is to fine-tune it, that is, to train it further on a few hundred to a few thousand reference calculations of the system at hand~\cite{grunert2025,atk,tompa2026fine,atk_size,haenseroth2026umlip_config_space}. 
In our own studies, foundation models of rather different architecture and size reached nearly the same error after fine-tuning~\cite{atk}. 
What a fine-tuned model inherits unchanged from its foundation is the cost of a force call.
Simulations limited by force calls and fine-tuned models thus ask for the same kind of model: one that balances zero-shot accuracy against speed rather than maximising the former.

Our fine-tuning studies~\cite{grunert2025,atk,atk_size,haenseroth2026umlip_config_space} make this the design principle of the present work: inference speed is as much a design target as zero-shot accuracy.
The pretrained model must be accurate and transferable enough across the chemical space to be used as it is, for example to sample the configurations on which the reference calculations for a fine-tune are made~\cite{haenseroth2026umlip_config_space}.
At the same time, it must keep enough capacity to reach close to ab initio accuracy for a specific family of compounds after a short fine-tune.
Benchmarks such as \textsc{Matbench} Discovery rank zero-shot accuracy, and most universal potentials have been developed towards it; how fast a model can be while meeting both requirements has received less attention.

Universal potentials designed for speed have emerged only recently, and they show that speed is a matter of design rather than of hardware. 
\textsc{NEP89}~\cite{nep89} does without message passing, in which every atom receives learned features that its neighbours have computed from their own environments, so that its energy depends on atoms beyond its cut-off; it runs in its own engine, \textsc{GPUMD}~\cite{gpumd}.
\textsc{DPA-4.1C}~\cite{dpa-4c} does without message passing as well and evaluates everything in one hand-written GPU kernel (Table~\ref{tab:compare}). 
Both tie their speed to a dedicated engine or kernel. 
Much everyday work, however, runs in general-purpose tools: \textsc{ASE}~\cite{ase}, \textsc{i-PI}~\cite{ipi3} and \textsc{LAMMPS}~\cite{lammps}.
There every step also pays for the tool around the model, and for a small model this part is not negligible: a thermostat in \textsc{ASE} or the construction of the neighbour list can cost more than the model itself (Sec.~\ref{sec:res-speed}).

\textsc{ILMlip} takes up the approach of these models and carries it into this setting.
Its place on the trade-off is a choice that we state at the outset (Fig.~\ref{fig:network}b).
It is not the most accurate universal potential. 
Its six sizes, from XS with $1.4\times10^5$ parameters to XXL with $2.8\times10^7$, reach F1 scores of $0.75$ to $0.91$ where the leader reaches $0.935$. 
What is gained depends on the comparison. 
Through \textsc{ASE} on one GPU, the small model S runs molecular dynamics (MD) of $10^4$~atoms $920$~times faster than the established message-passing potential \textsc{MACE-OMAT-0}~\cite{macemp0} (medium, $9.1$~million parameters), and three to five times faster than \textsc{NEP89} (in its \textsc{ASE} implementation) and \textsc{DPA-4.1C}. 
XL and XXL are the accuracy end of the family: still $104$ and $40$~times faster than \textsc{MACE-OMAT-0}, but slower than \textsc{NEP89} and \textsc{DPA-4.1C}.
Staying fast becomes hard only near the top of the accuracy scale, and that is not where \textsc{ILMlip} competes.

Almost every concept of \textsc{ILMlip} is known. 
The many-body products of the atomic cluster expansion~\cite{drautz2019,dusson2022} and of \textsc{MACE}~\cite{mace,mace_2} give it its body order; \textsc{DPA-4.1C} contributed pair-dependent modulations of the radial functions, fewer channels for higher angular momentum and a compact description of the last layer~\cite{dpa-4c}; \textsc{NEP89} the attitude that speed is a design goal from the first decision~\cite{nep89}.
Our contribution is the combination of these parts into one model, under a single rule for where a fast model may spend its work, and the parts that are new: element-dependent cut-offs; one pair object per bond, shared by both layers; a message without the part that came from the receiving atom; Gram matrices of a learned projection; the removal of many-body products that are numerical multiples of others; two smooth bounds that keep the features finite at any density and leave the shortest distances to a screened nuclear repulsion~\cite{zbl1985}; and the evaluation of many independent simulations in one network call.
Each of these parts can be adopted, replaced or extended on its own.

\begin{table*}[tp]
    \caption{\textbf{\textsc{ILMlip} in comparison with other fast universal potentials.} 
    The pretrained MLIPs \textsc{DPA-4.1C-Air-OAM}~\cite{dpa-4c} and \textsc{NEP89}~\cite{nep89} architectures are compared with the \textsc{ILMlip} models. 
    The \textsc{OMat24} errors are mean absolute errors on the \textsc{OMat24} validation set and taken for \textsc{NEP89} from the independent evaluation in Ref.~\onlinecite{dpa-4c}, with D3(BJ) added to the reference labels; for \textsc{DPA-4.1C} of its Air model trained on \textsc{OMat24} alone; for \textsc{ILMlip} of the OAM-OMAT models on a $2$~\% hold-out of \textsc{OMat24}. 
    Speed: molecular dynamics through \textsc{ASE} on one \textsc{Nvidia} B200, measured by us for all three (Sec.~\ref{sec:res-speed}).}
    \label{tab:compare}
    \setlength{\tabcolsep}{5pt}
    \renewcommand{\arraystretch}{1.3}
    \small
    \begin{tabular}{>{\raggedright}p{3.2cm}p{4.3cm}p{4.3cm}p{4.3cm}}
        \toprule
         & \textsc{NEP89}~\cite{nep89} & \textsc{DPA-4.1C-Air}~\cite{dpa-4c} & \textsc{ILMlip} (XS \ldots XXL) \\ 
        \midrule
        parameters & $9.8\times10^5$ & $4.3\times10^5$ & $1.4\times10^5$\ldots$2.8\times10^7$ \\
        learned features exchanged between atoms & none & none & one message (two layers) \\
        cut-off & $6$~{\AA} radial, $5$~{\AA} angular & $6$~{\AA} & $4$-$6$~{\AA} element-specific (XL, XXL: $6$~\AA) \\
        atomic neighbourhood range & $6$~\AA & $6$~\AA & $8$-$12$~{\AA} (XL, XXL: $12$~\AA) \\
        many-body terms & 3-, 4- and 5-body invariants ($35$~per atom) & Gram matrices, low-rank bispectrum and quartic (to body order 5) & ACE products to body order 4 (XXL: 3), projected Gram matrices, low-rank bispectrum \\
        chemistry & radial coefficients per ordered element pair & pair FiLM and mixing of radial profiles per ordered pair & group/period code, pair FiLM \\
        readout & one network per element & one network for all elements & one network for all elements, per-element bias and output vector \\
        short range & \textsc{ZBL} (1-2\,\AA) & none built in & ZBL with core switch \\
        training & evolution strategy, no gradients & Muon and Adam & Muon and AdamW \\
        engine & \textsc{GPUMD}; \textsc{LAMMPS} and \textsc{ASE} through interfaces & \textsc{DeePMD}-kit with a fused \textsc{CUDA} kernel, \textsc{LAMMPS} & \textsc{ASE}, \textsc{LAMMPS}; \textsc{JAX} with fused kernels \\
        \textsc{OMat24} val.\ $E$/$F$ MAE (meV/atom, meV/\AA) & $85.5$ / $269.9$ & $19.3$ / $105.4$ & $26.9 / 123$ (XS), $19.3$ / $101$ (L), $11.6$ / $67.3$ (XXL) \\
        \textsc{Matbench} Discovery F1 & -- & $0.799$ & $0.748$ (XS)\ldots$0.911$ (XXL) \\
        atom-steps/s, $10^4$ Cu atoms, B200, \textsc{ASE} & $2.2\times10^6$ & $3.7\times10^6$ & $1.2\times10^7$ (XS)\ldots$5.1\times10^5$ (XXL) \\
    \bottomrule
    \end{tabular}
\end{table*}

\section{The ILMlip model}
\label{sec:model}

\begin{figure*}[tp]
    \centering
    \includegraphics[width=\textwidth]{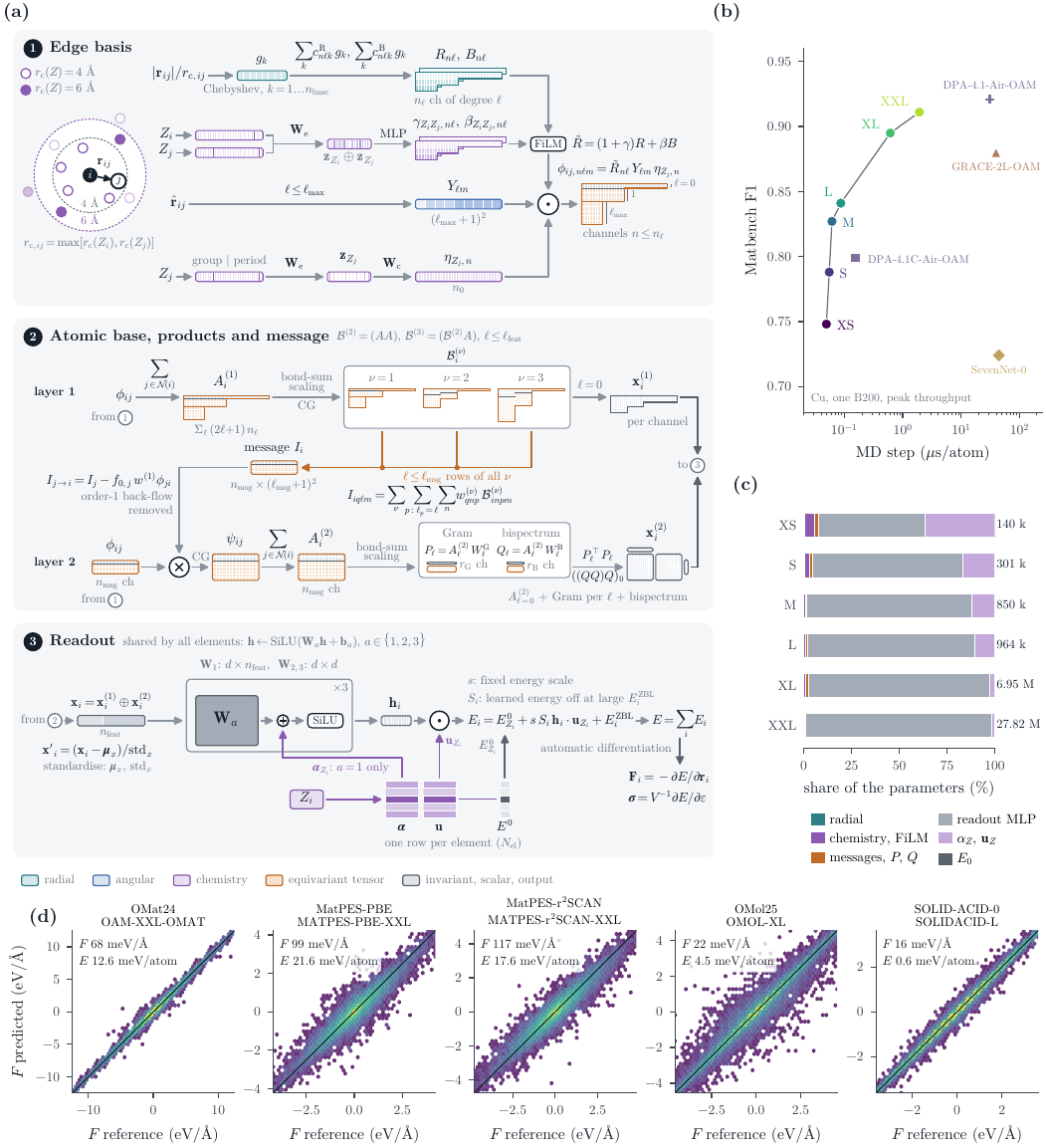}
    \caption{\textbf{Overview of the ILMlip models.}
    \textbf{(a)}~The three parts of the network. 
    1.~Each neighbour pair $i,j$ within its element cut-off $r_{\mathrm{c},ij}$ gets one pair object $\phi_{ij}$: Chebyshev radial functions modulated per element pair~\cite{perez2018film}, spherical harmonics and a chemistry factor of the neighbour.
    2.~The first layer sums these per atom, scales the sum (bond-sum gate) and forms products up to body order four; each neighbour's message, without the part from the receiving atom, builds the second layer. 
    3.~The invariants of both layers (second layer: scalars, Gram matrices, low-rank bispectrum) enter one readout network shared by all elements, with a bias and an output vector per element; a core switch removes the learned energy where the screened nuclear repulsion becomes large. 
    Symbols as in Sec.~\ref{sec:m-arch}. 
    \textbf{(b)}~MD-step cost per atom at peak throughput (fcc Cu, \textsc{ASE}, one B200) against the \textsc{Matbench} Discovery F1 score.
    \textbf{(c)}~Parameters of the OAM sizes and their share per part of the network.
    \textbf{(d)}~Force parity on held-out structures of OAM-XXL-OMAT, the \textsc{MatPES} XXL models, OMOL-XL and SOLIDACID-L.}
    \label{fig:network}
\end{figure*}

A force call of a message-passing potential does two kinds of work. 
Work per bond, that is per pair of neighbouring atoms, is repeated about $20$ to $80$~times per atom (typical numbers of neighbours within $6$~\AA) and moves data through memory. 
Work per atom is done once and can be arranged as dense matrix products, which GPUs and CPUs execute with high performance. 
To see which of the two sets the cost, we timed one call of a randomly initialised model and varied one hyperparameter at a time: the cut-off radius, the angular momentum of what travels along a bond and the number of layers set the time, while more channels per atom and a higher body order cost a few per cent and a wider readout network was cheap per bond but not free (Supplementary Note~5).
\textsc{ILMlip} is built on one rule: spend little per bond and as much as needed per atom (Fig.~\ref{fig:network}a).

Three decisions keep the bond cheap.
The first is the atomic cluster expansion~\cite{drautz2019}, which sums simple functions of each neighbour's position into a per-atom base and only then forms many-body products from it.
Building the base takes work per bond; the products, three base functions for body order four, are formed once per atom, so their cost does not grow with the number of neighbours.
The second is the cut-off, the distance within which two atoms count as neighbours.
One radius for all elements would have to fit the largest ions and would hand every compact metal several times the neighbours it needs.
Each element therefore gets its own cut-off between $4$ and $6$~{\AA} according to its size, and a pair uses the larger of the two: copper (fcc) keeps $18$~neighbours instead of the $78$~within $6$~\AA.
The price is a direct range of only $4$~{\AA} between elements such as H, C, O and Cu; what lies beyond reaches an atom only through the message described below.
The third decision is that each bond carries a single object, built once and used by both layers: radial functions for its length times spherical harmonics for its direction.
A feature-wise linear modulation (FiLM)~\cite{perez2018film,dpa-4c} adapts the radial part to the pair of elements.
Since this modulation depends only on the two elements, it is a fixed table at inference time, and the chemistry of a pair costs a table lookup, not a network evaluation.

On the side of the atom, \textsc{ILMlip} can afford what the bond cannot. 
The first layer forms all channel-wise products of its base up to body order four. 
In the second layer every neighbour sends a message, a learned summary of its own surroundings, which doubles the range of an atom's energy to two cut-offs. 
\textsc{NEP89} and \textsc{DPA4C} exchange no learned features, \textsc{ILMlip} pays for its message there.
To keep the payment small, this one message is the only exchange, and the second layer forms no products of its own.
It also differs from the layers of message-passing potentials such as \textsc{MACE}~\cite{mace}, which couple neighbour features and bond directions in tensor products whose weights a radial network computes for every bond and coupling path: \textsc{ILMlip} couples each channel of the message with the same channel of the pair object by fixed Clebsch-Gordan coefficients, so that the bond adds no learned weights of its own.
Its invariants, the quantities that do not change when the structure is rotated, are Gram matrices (scalar products between channels) and a low-rank bispectrum (a reduced set of triple products), following \textsc{DPA4C}~\cite{dpa-4c}, which takes the Gram matrices of a full-width channel map; \textsc{ILMlip} additionally takes them of a learned projection to $r_{\mathrm{G}}$ columns.
The bond sum of the second layer holds several coupling paths per message channel, and the Gram matrix of all of them would grow with the square of their number; that of a projection to $r_{\mathrm{G}}$ columns has $r_{\mathrm{G}}(r_{\mathrm{G}}+1)/2$ entries per degree, whatever the number of paths.
All invariants enter one readout network shared by all elements.
It holds between $56$~\% (XS) and $97$~\% (XXL) of the parameters (Fig.~\ref{fig:network}c).

A fast model makes long simulations affordable, and these visit configurations that no training set contained.
Two problems arise: many-body products grow polynomially with the local density, and at very short distances a model can learn spurious wells. 
Against the first, \textsc{ILMlip} scales every bond sum $A_i$, the sum of the pair objects over the neighbours of atom $i$, in each of the two layers by a smooth per-atom factor ($f_{0,i}$ and $f_{1,i}$ in Sec.~\ref{sec:m-arch}, each with its own $\sigma$),
\begin{equation}
    f_i=\frac{a_0}{\sigma}\Bigl(1+\frac{\rho_i^2}{\kappa^2}\Bigr)^{-1/2}, \qquad \rho_i^2=\frac{\langle A_i^2\rangle}{\sigma^2},
    \label{eq:gate}
\end{equation}
where $\langle A_i^2\rangle$ is the mean square over the channels and components of the bond sum, $\sigma$ its root mean square in the freshly initialised network on the first training batches (kept fixed afterwards), and $a_0=0.05$ sets the scale seen by the products. 
Whatever the density, the root-mean-square component of $f_iA_i$ cannot exceed $a_0\kappa$, while at the chosen $\kappa=16$ a bond sum of the calibration magnitude ($\rho_i=1$) is reduced by only $0.2$~\%. 
In the released models the bond sums stay below this magnitude rather than above it: on held-out structures of their own training data the median $\rho_i$ is $0.4$ in the first layer of S to L, $0.7$ in XS and $0.06$ to $0.1$ in XL and XXL, and at most $0.2$ in the second layer, and $\rho_i>1$ occurs for $9$ to $11$~\% of the atoms of XS, for at most $0.1$~\% of those of S and M and for none of the others (Supplementary Note~4). 
The gate therefore leaves the median atom within $0.1$~\% of its low-density scaling and acts only on atoms whose bond sum is far above that of the training data; at $\rho_i=\kappa$ the factor is $0.71$.

Against the second problem, the learned energy of an atom is multiplied by a switch $S_i$ that goes smoothly from one to zero as the atom's screened nuclear repulsion: a screened nuclear repulsion (Ziegler-Biersack-Littmark model, ZBL)~\cite{zbl1985} rises from zero to $10$~eV, so that inside this core only the physical repulsion acts. 
Both bounds are per-atom operations. 
In trainings of S and L on \textsc{MatPES}-PBE with both, with either and with neither, the test errors differ by at most $0.6$~meV/atom ($1.8$~\%) in energy and $1.5$~meV/{\AA} ($1.3$~\%) in force, where two seeds of the same recipe differ by $0.4$ to $0.6$~meV/atom and $0.7$ to $1.1$~meV/{\AA}; the model with both bounds is $0.1$~\% worse in energy and $0.6$~\% in force at both sizes than the one with neither (Supplementary Note~3).
Both bounds were introduced after development versions of the architecture without them, driven far outside the training distribution in stress tests, showed features that grew without limit at high local density and spurious attractive wells at very short interatomic distances.

\section{Results}
\label{sec:results}

\begin{figure*}[tp]
    \centering
    \includegraphics[width=\textwidth]{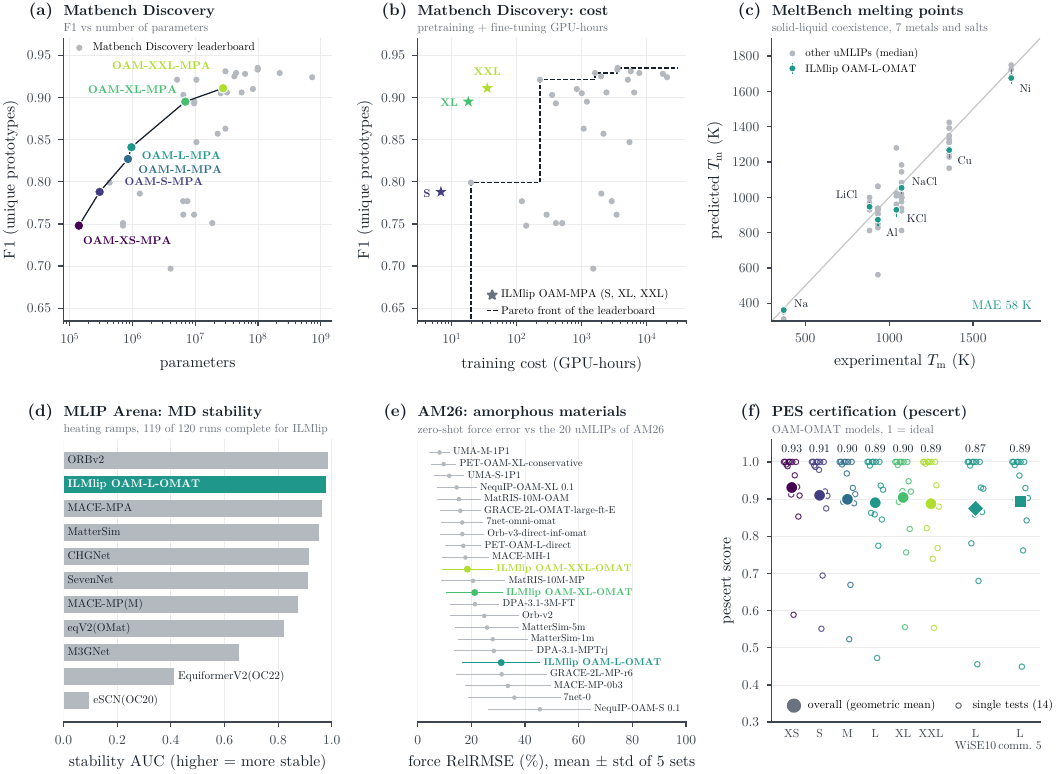}
    \caption{\textbf{Benchmarks of the \textsc{ILMlip} OAM models.}
    \textbf{(a)}~\textsc{Matbench} Discovery F1 score~\cite{matbench} on the unique prototypes against the number of parameters for the OAM-MPA models (pretrained on \textsc{OMat24}, \textsc{sAlex} and \textsc{MPtrj}, fine-tuned on
    \textsc{MPtrj} and \textsc{sAlex}) and $30$~models of the \textsc{Matbench} Discovery leaderboard with a reported training cost (grey); the submitted OAM-S-MPA was pretrained for 8 instead of 30 epochs.
    \textbf{(b)}~The same F1 scores against the training cost in GPU-hours (pretraining plus fine-tuning, as reported for each model on the leaderboard) with the current Pareto front of the leaderboard (dashed); OAM-S-MPA, OAM-XL-MPA and OAM-XXL-MPA took 6.9, 18.1 and 35.2 GPU-hours (\textsc{Nvidia} H100 NVL/B200).
    \textbf{(c)}~Melting points of seven metals and salts from solid-liquid coexistence simulations with OAM-L-OMAT against experiment; grey: other universal potentials from the job archive of \textsc{MeltBench}~\cite{meltbench} (median of the repeated runs of each model); error bars: the temperature interval in which the coexisting phases switch from crystal growth to melting. 
    \textbf{(d)}~MD stability in the \textsc{MLIPArena} heating benchmark~\cite{mliparena}: area under the curve of the fraction of completed simulations, OAM-L-OMAT against the models reported by \textsc{MLIPArena}. 
    \textbf{(e)}~Zero-shot force errors on the \textsc{AM26} benchmark~\cite{fragapane2026am26} of amorphous materials (a-C, a-Si, a-SiO$_2$, a-GST, a-LiPS): relative force RMSE, mean $\pm$ standard deviation over the five sets, for OAM-L/XL/XXL-OMAT and the $20$~universal potentials of the \textsc{AM26} reference.
    \textbf{(f)}~Potential energy surface consistency certification by \textsc{pescert}~\cite{haenseroth2026pescert} of the OAM-OMAT models, the weight average of ten second-stage runs of OAM-L-OMAT (WiSE10, Sec.~\ref{sec:m-ensembles}) and a committee of five OAM-L-OMAT models: overall score (filled; geometric mean of $14$~tests) and the single tests (open); 1 = ideal.}
    \label{fig:benchmarks}
\end{figure*}

\begin{figure*}[tp]
    \centering
    \includegraphics[width=\textwidth]{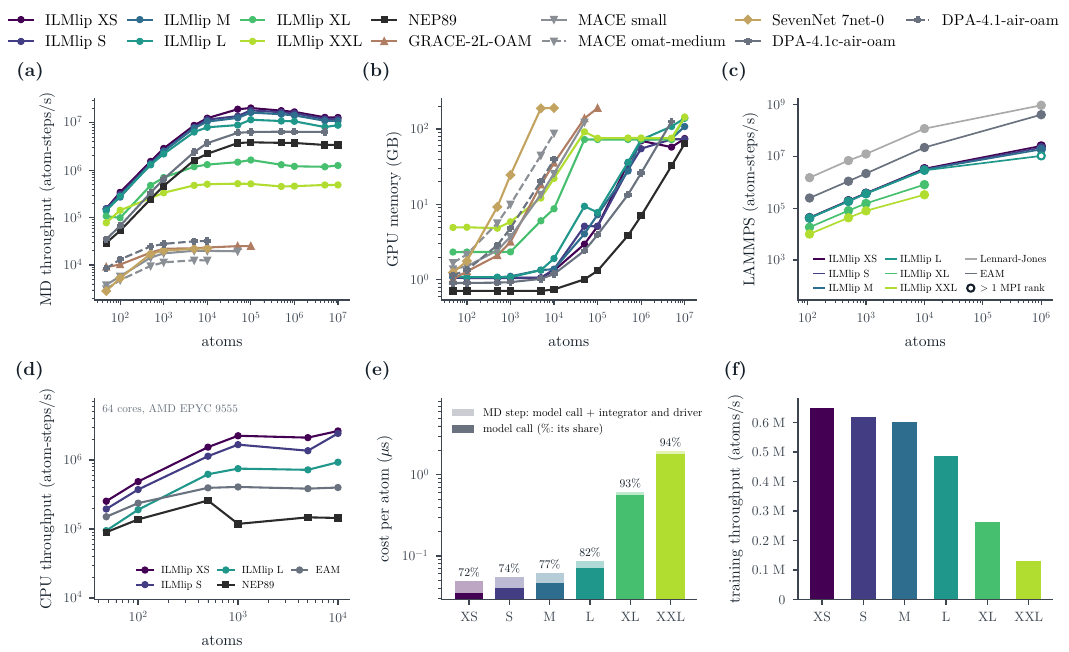}
    \caption{\textbf{Throughput and timing benchmarks.}
    All tests were run on fcc Cu, $300$~K, $1$~fs, with the \textsc{MatPES}-PBE weights of all sizes, on one \textsc{Nvidia} B200 unless stated. 
    \textbf{(a)}~MD throughput through \textsc{ASE} against the number of atoms for \textsc{ILMlip} XS-XXL (\textsc{JAX}, fused kernels) and published universal potentials; \textsc{NEP89} run to $2\times10^7$ atoms, a curve of another model ends at the largest size it ran (XS-M ran $10^8$ atoms, L $7\times10^7$ atoms, XL and XXL $2\times10^7$ atoms on one \textsc{Nvidia} B200, Supplementary Note~5). 
    \textbf{(b)}~GPU memory of the same runs, including the pool that \textsc{JAX} reserves, so that the plateau of XL and XXL near $75$~GB is not the need of the model. 
    \textbf{(c)}~MD throughput in \textsc{LAMMPS} (ML-IAP~\cite{lammps} with KOKKOS~\cite{lammps_kokkos}) next to Lennard-Jones and EAM; open symbols: several MPI ranks share the GPU. 
    \textbf{(d)}~CPU throughput of the CPP engine (XS, S, L) on the $64$~cores of an \textsc{AMD EPYC} 9555, next to \textsc{NEP89} (\textsc{ASE}) and EAM (\textsc{ASE}, numpy) on the same cores.
    \textbf{(e)}~Cost of one MD step per atom at $10^5$~atoms (light) and the part of it spent in the model call, the network evaluated on the GPU including the transfers between host and GPU (dark); the rest is the integrator and the driver.
    \textbf{(f)}~Training throughput with the production recipe (NaCl, $64$-atom cells, TF32).}
    \label{fig:speed}
\end{figure*}

\subsection{Accuracy over a range of benchmarks}
\label{sec:res-price}

Figure~\ref{fig:benchmarks} combines \textsc{Matbench} Discovery, by which the field ranks universal potentials, with tests away from relaxed crystals and with a check of the potential energy surface itself; errors on held-out structures of every training set are in Fig.~\ref{fig:network}d, Table~\ref{tab:compare} and Supplementary Note~6.

On \textsc{Matbench} Discovery the models behave as the trade-off promises.
The F1 score of the MPA models rises with size from $0.75$ for XS to $0.91$ for XXL, $80$ to $97$~\% of the leader's $0.935$ (Fig.~\ref{fig:benchmarks}a).
The submitted S, with $0.30$~million parameters, reaches $0.788$ next to the $0.799$ of \textsc{DPA-4.1C-Air} with $0.43$ million, a model that is also built for speed; XL passes \textsc{GRACE-2L-OAM}, $0.895$ against $0.880$, with $55$~\% of its parameters. 
The score rates how well energies rank crystals, and what a small model gives up appears where second derivatives of the energy enter. 
Phonons, and with them the thermal conductivity, are the clearest weakness, at least of the small sizes: the symmetric relative error $\kappa_\mathrm{SRME}$ ($0$ is perfect) is $0.63$ for S and $0.22$ for XXL, against $0.08$ for the current leader (\textsc{DPA-4.1-Max-OAM-PFT}).
Models widely used for molecular dynamics lie in between, \textsc{MACE-MPA-0} (medium)~\cite{macemp0} at $0.412$ and \textsc{GRACE-2L-OAM-L}~\cite{grace_2} at $0.294$, so XXL is already more accurate here than both.
S and XXL took $6.9$ and $35$~GPU-hours from scratch (including the second stage on the MPA energy scale), which brackets the $20$~GPU-hours of \textsc{DPA-4.1C-Air} and is $100$ to $500$~times less than the $3.6$~thousand GPU-hours of the current leader, \textsc{DPA-4.1-Max-OAM-PFT}.
No published model reaches the F1 score of S, XL or XXL with less training: all three lie to the left of the Pareto front that the leaderboard draws in F1 against training cost (Fig.~\ref{fig:benchmarks}b).

Besides relaxed crystals the models are well behaved and sit in the middle of the field.
Solid-liquid coexistence simulations with OAM-L-OMAT give the melting points of seven metals and salts with a mean absolute error of $58$~K, inside the spread of the universal potentials in the job archive of \textsc{MeltBench}~\cite{meltbench} (Fig.~\ref{fig:benchmarks}c).
In the heating test of \textsc{MLIPArena}~\cite{mliparena}, $119$ of $120$~runs complete, second only to \textsc{ORBv2} among the ten models reported there (Fig.~\ref{fig:benchmarks}d). 
Amorphous structures are harder in general: on \textsc{AM26}~\cite{fragapane2026am26}, the force error falls from $31$~\% of the force spread for L to $19$~\% for XXL, which places L, XL and XXL at ranks $17$, $12$ and $11$ of $21$ when each is placed among the $20$ published models (Fig.~\ref{fig:benchmarks}e).
The clearest failure is in molten alkali chlorides, whose densities come out $11$ to $18$~\% too low (Supplementary Note~8). 
Part of this error is inherited from the reference: \textsc{OMat24} is computed with PBE without a dispersion correction, and PBE without dispersion is known to underestimate the density of molten alkali halides considerably~\cite{bengtson2014}. 

A different kind of test asks whether the potential energy surface is physical at all. 
The suite \textsc{pescert}~\cite{haenseroth2026pescert} checks fourteen properties that any such surface has, such as energy conservation, equivariance and smoothness. 
The six OMAT models reach overall scores between $0.89$ and $0.93$ (Fig.~\ref{fig:benchmarks}f).
We also release weight averages~\cite{wortsman2022soups} of ten second-stage runs (WiSE10, Sec.~\ref{sec:m-ensembles}) of L and XXL and committees of five models of L.
On \textsc{Matbench} Discovery the WiSE10 average of OAM-L-MPA scores like the single model, F1 $0.842$ against $0.841$, while the committee of five OAM-L-MPA models gains $0.009$ in F1 at five times the cost of every step.
The WiSE10 average of OAM-L-OMAT fails the stress consistency test of \textsc{pescert}, which the single model passes (Supplementary Note~6).

\subsection{Speed in general-purpose tools}
\label{sec:res-speed}

Speed is measured as the throughput of MD in atom-steps per second, for fcc copper on one \textsc{Nvidia} B200 through \textsc{ASE} (Fig.~\ref{fig:speed}a). 
For $10^4$ atoms, \textsc{ILMlip}-S reaches $1.16\times10^7$~atom-steps/s, so a nanosecond of this cell at a $1$~fs time step takes less than $15$~minutes.
The two fast models in this benchmark, \textsc{NEP89} and \textsc{DPA-4.1C-Air}, reach $2.2\times10^6$ and $3.7\times10^6$~atom-steps/s on the same card and in the same driver, so S is $5.3$ and $3.1$~times faster. 
The message-passing potentials that we have run, \textsc{MACE-MP-0} (small), \textsc{MACE-OMAT-0} (medium), \textsc{SevenNet-0}, \textsc{GRACE-2L-OAM} and \textsc{DPA-4.1-Air}, reach $1.3$ to $3.2\times10^4$~atom-steps/s, $360$ to $920$~times less on the same GPU and system. 
One of the most widely used of them, \textsc{MACE-OMAT-0} (medium, same architecture as \textsc{MACE-MPA-0}, F1 $0.852$) reaches $1.26\times10^4$~atom-steps/s, $920$~times less than S. 
The gap lies in the model itself: one force call on $10^4$ atoms, without integrator and neighbour list, costs $45$~$\mu$s per atom for \textsc{MACE-OMAT-0} and $0.063$~$\mu$s for S, $710$~times less, and at $5\times10^4$~atoms \textsc{MACE-OMAT-0} no longer fits into the GPU, where S needs $2.1$~GB at $10^5$~atoms (Supplementary Note~5).

On one GPU (\textsc{Nvidia} B200, $183$~GB), XS-M run $10^8$~atoms, L runs $7\times10^7$, XL and XXL run $2\times10^7$ (Supplementary Note~5). 

Within the family from XS to L, one model call at $10^5$ atoms rises only from $35$ to $71$~ns/atom, because L spends its additional parameters per atom and bond. 
XL and XXL cost $0.56$ and $1.8$~$\mu$s: they use a $6$~{\AA} cut-off for every element, which raises the number of bonds of copper from $18$ to $78$, and carry higher angular momenta along each bond (Fig.~\ref{fig:speed}e). 
At $10^4$~atoms XL therefore runs at $0.59$ of the throughput of \textsc{NEP89} and XXL at $0.23$, and both are slower than \textsc{DPA-4.1C-Air} as well (Supplementary Note~5), though still $104$ and $40$~times faster than \textsc{MACE-OMAT-0} in MD, and $76$ and $27$~times in the force call alone.

For models this cheap, the engine around the model matters. 
In plain velocity-Verlet dynamics the model call takes $72$ to $94$~\% of a step, depending on the size (Fig.~\ref{fig:speed}e); for S and $10^5$~atoms it is $4.1$ of $5.5$~ms. 
A thermostat changes the balance. 
With \textsc{ASE}'s Nos\'e-Hoover chain the same step takes $9.8$~ms and with \textsc{ASE}'s Langevin thermostat $16$ to $22$~ms, so that the integrator then costs more than the model and deserves as much care in the choice as the model size. 
The neighbour list is the other place where the balance can tip: rebuilt at every step instead of only when atoms have moved by part of a skin, it makes S $30$ to $500$~times slower, depending on the code that builds it (Supplementary Note~5).

The speed is not tied to \textsc{ASE}, to GPUs or to large cells.
In \textsc{LAMMPS}, \textsc{ILMlip}-S runs $10^6$~atoms at $2.2\times10^7$~atom-steps/s (Fig.~\ref{fig:speed}c), only $19$~times slower than the embedded-atom method (EAM), a classical potential for metals, on the same GPU.
Without a GPU, the CPP engine on the $64$~cores of a CPU node runs $10^4$~atoms at $2.4\times10^6$~atom-steps/s, $17$~times \textsc{NEP89}'s CPU calculator (\textsc{ASE}) in the same setting (Fig.~\ref{fig:speed}d), and just one core of a conventional laptop reaches $5.2\times10^4$~atom-steps/s with S at $10^4$~atoms. 
A single small cell leaves most of a GPU idle, and many cells in one call put it to use: for S, $32$~cells of $96$~atoms in one call are evaluated $11$~times faster than one call per cell, the $32$~beads of a path-integral simulation $5.6$~times faster, and $64$~molecular-dynamics runs in lockstep run $6.8$~times faster than one after another, where $64$~separate processes sharing the GPU gain only a factor of $1.2$ (Sec.~\ref{sec:m-batched}; Supplementary Note~5). 
Training profits from the same design (Fig.~\ref{fig:speed}f): the OAM pretraining ($8$~epochs) of the submitted S took $6.2$~GPU-hours.

\textsc{DPA-4.1C-Air} runs there through the calculator of \textsc{DeePMD}-kit~\cite{deepmd_v3} itself, whereas \textsc{NEP89} was built for its own engine, \textsc{GPUMD}~\cite{gpumd}, where it was reported at $7.0\times10^6$~atom-steps/s for copper on a V100, three times the $2.2\times10^6$ that we measure for it through \textsc{ASE} on the B200; we did not run \textsc{GPUMD}.

\begin{figure*}[tp]
    \centering
    \includegraphics[width=\textwidth]{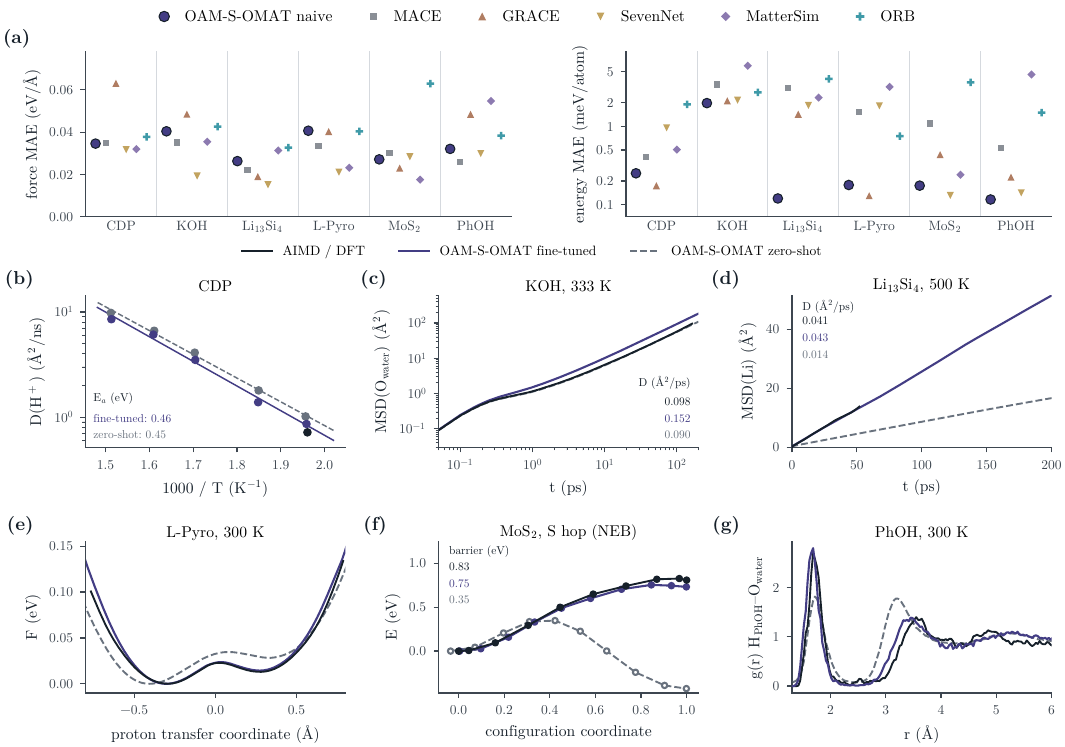}
    \caption{\textbf{Fine-tuning performance of ILMlip.}
    \textbf{(a)}~Force and energy MAE on the test trajectories of six systems after naive fine-tuning of OAM-S-OMAT, next to the fine-tuned foundation models of Ref.~\onlinecite{atk} (\textsc{MACE-MP-0}, \textsc{GRACE-1L-OAM}, \textsc{SevenNet-0}, \textsc{MatterSim-v1.0.0-5M}, \textsc{ORBv2}).
    \textbf{(b-g)}~Production MD of the fine-tuned (solid) and the zero-shot (dashed) \textsc{ILMlip} OAM-S-OMAT against AIMD/DFT (black):
    \textbf{(b)}~Arrhenius plot of the proton diffusion coefficient in CsH$_2$PO$_4$ with the activation energies; 
    \textbf{(c)}~mean squared displacement (MSD) of the water oxygen in aqueous KOH at $333$~K with the diffusion coefficients; 
    \textbf{(d)}~Li-ion MSD in Li$_{13}$Si$_4$ at $500$~K;
    \textbf{(e)}~free-energy profile of the proton transfer in L-pyroglutamic acid at $300$~K; 
    \textbf{(f)}~NEB profile of the sulphur hop in defective 2D MoS$_2$ with the barriers (zero-shot: the maximum of a path that ends below its start); 
    \textbf{(g)}~radial distribution function of the hydroxyl hydrogen of phenol and the water oxygen at $300$~K.}
    \label{fig:finetune}
\end{figure*}

\subsection{Fine-tuned models are on par with well-established message-passing models}
\label{sec:res-ft}

The second use of a fast model, as the starting point of a fine-tune, rests on a claim that can be tested: after fine-tuning, a small model should reach the errors of large ones. 
We fine-tuned OAM-S-OMAT on six materials of our earlier study~\cite{atk}, from a solid acid~\cite{struct_cdp,dressler2020effect,dressler2023coexistence} to a Li-Si alloy\cite{zeilinger2013revision,kirsch2022atomistic,qaisrani2025bridging}, an organic crystal~\cite{qaisrani2025acid,haenseroth2026nqe_lpyro}, an aqueous phenol and potassium hydroxide solution~\cite{hanseroth2025optimizing}, each with about $2{,}000$~frames of ab initio molecular dynamics (AIMD) and with one recipe for all (Sec.~\ref{sec:m-ft}). 
Each fine-tune trained for $3$ to $16$~minutes on one GPU (\textsc{Nvidia} B200 or H200 NVL).

Fine-tuning lowers the force error of the zero-shot model by a factor of $4$ to $13$. 
Averaged over the six systems, the fine-tuned S has a force error of $33.5$~meV/{\AA} where the fine-tuned \textsc{MACE} of Ref.~\onlinecite{atk} has $30.3$~meV/{\AA}, it lies inside the range that the fine-tuned \textsc{SevenNet}, \textsc{MatterSim}, \textsc{GRACE} and \textsc{ORB} span there, $24.3$ to $42.5$~meV/{\AA}; its energy error of $0.47$~meV/atom is below the $1.68$ of \textsc{MACE} (Fig.~\ref{fig:finetune}a; all six systems in Supplementary Note~7). 
A control puts the pretraining in perspective.
The same architecture trained from random weights on the same frames reaches the same force error, $33.3$~meV/{\AA}, and adapters (LoRA) or a second output head give higher ones (Supplementary Note~7). 

Low force errors do not guarantee correct observables, and a fine-tune is worth its cost only if it changes them. 
The fine-tuned and the zero-shot S model were therefore each run for up to $3$~ns per system at constant temperature, and the observables were compared with the ab initio reference (Fig.~\ref{fig:finetune}b-g). 
They fall into two groups, according to what the zero-shot model gets wrong.

The first group is what the zero-shot model gets clearly wrong and the fine-tune repairs. 
The Li diffusion coefficient in Li$_{13}$Si$_4$ becomes $0.043$~{\AA}$^2$/ps against $0.041$ from AIMD, where the zero-shot model gives $0.014$. 
The barrier of the sulphur jump in MoS$_2$ becomes $0.75$~eV against $0.83$~eV in the reference, whereas the zero-shot path has a maximum of $0.35$~eV and then falls $0.43$~eV below its start, a downhill path without the ab initio barrier shape. 
The height of the first peak of the phenol-water radial distribution function (near $1.8$~{\AA}) is $2.80$ against $2.70$ from AIMD, where the zero-shot model gives $1.82$.

The second group is what the zero-shot model already gets right, so that the fine-tune has little to add. 
The free-energy profile of the proton transfer in L-pyroglutamic acid is close to the ab initio one for both models. 
For proton diffusion in CDP, the activation energy is $0.46$~eV fine-tuned and $0.45$~eV zero-shot, both close to the $0.4$~eV of the literature~\cite{wang_superprotonic}, and the diffusion coefficient at $510$~K is within a factor of $1.4$ of AIMD for the zero-shot model and of $1.2$ for the fine-tuned (Supplementary Note~7).
In aqueous KOH the zero-shot model gives the diffusion coefficient of the water about $8$~\% below that of AIMD, $0.090$ against $0.098$~\AA$^2$/ps. 
The fine-tune lowers the force error on this system from $152$ to $40$~meV/{\AA} and makes the diffusion $1.6$~times too fast ($0.152$~\AA$^2$/ps).

The speed carries over, within the limit that small cells set: on these cells of $144$ to $512$~atoms the fine-tuned S ran at $0.6$ to $2.3\times10^5$~atom-steps/s, below its peak because a single small cell leaves most of the GPU idle. 
At about $450$~steps per second on one B200, $3$~ns of $0.5$~fs steps take $3.7$~hours.

\section{Discussion}
\label{sec:discussion}

The results locate \textsc{ILMlip} on the trade-off it was built for: fast enough to make large and long simulations routine, accurate enough to be used zero-shot across chemistry, and transferable enough that a cheap fine-tune on little data brings reference accuracy to this speed.

The sizes XS to L reach $80$ to $90$~\% of the leading F1 score, $0.75$ to $0.84$, a range in which the leaderboard holds models of about one to ten million parameters, such as \textsc{SevenNet-l3i5} ($0.76$ with $1.2$~million) and \textsc{MACE-MPA-0} ($0.85$ with $9.1$~million); XS to L have $0.14$ to $0.96$~million parameters. 
In \textsc{ASE} a step of XS costs $990$~times less than one of \textsc{MACE-OMAT-0} (medium) and its force call $800$~times less; for XL and XXL the factors shrink to $104$ and $40$.
This suggests a division of labour by task rather than a single best model. 
Screening, large cells and long trajectories are limited by the number of force calls, and there S and M are the natural choice, S also on a laptop. 
When a property must be trusted, the accuracy comes from a fine-tune, which S supports at its own speed. 
Its force error after fine-tuning lies between those of the fine-tuned \textsc{MACE}, \textsc{SevenNet} and \textsc{MatterSim}, which are lower, and those of \textsc{ORB} and \textsc{GRACE}, which are higher, and its energy error is the lowest of the six. 
XL and XXL suit tasks in which zero-shot accuracy weighs slightly more than throughput, and they remain considerably faster than established models of similar F1 score.

The choices that make \textsc{ILMlip} fast also define the tasks for which it is not the most suitable potential.
When the highest zero-shot accuracy currently achievable is the objective, as in the discovery of stable crystals without a budget for ab initio calculations, the largest and most expensive universal potentials may be worth their cost.
Interactions beyond two cut-offs, such as long-range electrostatics at charged interfaces, are not represented.
Outside these cases, \textsc{ILMlip} is a well-founded choice: the accuracy of universal potentials is at present limited more by the quality of their training data than by their architecture, so speed decides, and a short fine-tune on high-quality data closes most of the remaining gap to the ab initio reference.

Against the two fast models \textsc{NEP89} and \textsc{DPA-4.1C-Air}, which also run in general-purpose tools, \textsc{ILMlip} pays for one message between atoms and is still faster there: through \textsc{ASE}, S is three to five times faster than both. 
The message doubles the range of an atom's energy to $8$ to $12$~{\AA}, and it is the reason why \textsc{ILMlip} needs ghost atoms of two cut-offs in parallel GPU runs with \textsc{LAMMPS} where the other two exchange only positions.
Its advantage is therefore speed inside the tools that most simulations already use, with their integrators, thermostats, enhanced-sampling and path-integral methods and analysis, so that its speed reaches existing workflows without porting them to a dedicated engine.

We expect future models to improve on \textsc{ILMlip}.
Its contribution is to show that the region of the trade-off set out in the introduction can be reached by combining design principles from different architectures under one cost rule, with new components added where the rule demands them.
Each component can also be adopted on its own, as a starting point for architectures with still better ratios of speed to accuracy.

We release the models of all sizes and training sets together with the code, the CPP, \textsc{JAX}, \textsc{LAMMPS} and \textsc{i-PI} interfaces and the fine-tuning tools, as a general-purpose potential for fast molecular dynamics and high-throughput simulations.

\section{Methods}
\label{sec:methods}

\subsection{Architecture}
\label{sec:m-arch}

Atom $i$ at $\mathbf r_i$ with atomic number $Z_i$ has as neighbours all atoms $j$ with $r_{ij}=|\mathbf r_j-\mathbf r_i+\mathbf S_{ij}|<r_{\mathrm c,ij}$, where $\mathbf S_{ij}$ is a lattice translation. 
Every element has one cut-off $r_{\mathrm c}(Z)$, which depends only on the element, and a pair uses the larger of the two, $r_{\mathrm c,ij}=\max[r_{\mathrm c}(Z_i),r_{\mathrm c}(Z_j)]$. 
For the elements that typically occur as anions, N, O, F, P, S, Cl, As, Se, Br, Sb, Te and I, $r_{\mathrm c}(Z)$ is $4$~{\AA} scaled by the ratio of their ionic radius to that of O$^{2-}$, so that two anions in contact lie within each other's cut-off. 
For all other elements it is twice the sum of the element's covalent radius and that of oxygen, so that the first shell of an oxide lies inside the smooth part of the cut-off function. 
The values are clipped to $4$-$6$~{\AA}, which gives $4$~{\AA} for $21$~elements, among them H, C, O, Si, Fe and Cu, and $6$~{\AA} for P, As, Sb, Te, I, Cs and Fr. 
XL and XXL use $6$~{\AA} for every pair, because at these sizes the cost per bond is small compared with that of the readout.

With $t=r_{ij}/r_{\mathrm c,ij}$, the radial basis consists of Chebyshev polynomials, as in \textsc{NEP89}~\cite{nep89}, times a polynomial envelope~\cite{gasteiger2020}, $g_k=T_k(2t-1)\,e(t)$, $e(t)=1-21t^5+35t^6-15t^7$ for $t<1$, $k=1\ldots n_{\mathrm{base}}$, which vanishes with two derivatives at the cut-off. 
Two sets of learned linear combinations, $R_{n\ell}=\sum_k c^{\mathrm R}_{n\ell k}g_k$ and $B_{n\ell}=\sum_k c^{\mathrm B}_{n\ell k}g_k$, give the radial functions of channel $n$ and degree $\ell\le\ell_{\max}$. 
Channel counts decrease with degree, $n_0\ge n_1\ge\ldots$, because scalars carry most of the radial and chemical information while high-degree channels dominate the cost of products. 
An element is described by a two-hot code of its group and period~\cite{twohotencoding1,twohotencoding2}, mapped by a learned matrix $\mathbf W_{\mathrm e}$ to a vector $\mathbf z_Z$, so that elements of one group or period share parameters. 
For the ordered pair of elements of a bond, a small network maps $\mathbf z_{Z_i}\oplus\mathbf z_{Z_j}$ to modulations $\gamma$ and $\beta$~\cite{perez2018film,dpa-4c}, and 
\begin{equation}
    \tilde R_{n\ell}=(1+\gamma_{Z_iZ_j,n\ell})R_{n\ell}+\beta_{Z_iZ_j,n\ell}B_{n\ell}.
    \label{eq:film}
\end{equation}
The modulation is initialised to zero and forms one table per pair at inference time; without it, the errors on \textsc{MatPES} rose by about twice their run-to-run scatter (Supplementary Note~3). 
The pair object is
\begin{equation}
    \phi_{ij,n\ell m}=\tilde R_{n\ell}(r_{ij})\,Y_{\ell m}(\hat{\mathbf r}_{ij})\,\eta_{Z_j,n},
    \label{eq:phi}
\end{equation}
with real spherical harmonics $Y_{\ell m}$ and a chemistry factor $\eta_{Z_j}=\mathbf W_{\mathrm c}\mathbf z_{Z_j}$ of the neighbour.

The atomic base is the sum over neighbours $j$, $A^{(1)}_{i,n\ell m}=\sum_{j}\phi_{ij,n\ell m}$, scaled by the factor $f_{0,i}$ of Eq.~\eqref{eq:gate}. Products within each channel,
\begin{equation}
    \mathcal B^{(1)}=A,\quad\mathcal B^{(2)}=(A\otimes A),\quad \mathcal B^{(3)}=(\mathcal B^{(2)}\otimes A),
    \label{eq:cg_products}
\end{equation}
are coupled with real Clebsch-Gordan coefficients to degrees $L\le\ell_{\mathrm{feat}}$; only the pair products see degrees above $\ell_{\mathrm{feat}}$, up to $\ell_{\max}$, because high degrees contribute little to higher body orders in the cluster expansion~\cite{drautz2019}. 
Couplings of order three that are numerical multiples of another are dropped, about a fifth of the products. 
The invariants of the first layer, $\mathbf x^{(1)}_i$, are all products with $L=0$ of all channels and orders. 
Body order runs to four ($\nu\le3$ factors of $A$) in XS to XL.
XXL stops at $\nu=2$ and uses its budget for features and messages up to degree $\ell=3$ instead ($\ell_{\mathrm{feat}}=\ell_{\mathrm{msg}}=3$, Supplementary Note~1).

Every atom forms a message, a linear map of its products of degree $\ell\le\ell_{\mathrm{msg}}$ to $n_{\mathrm{msg}}$ channels,
\begin{equation}
    I_{i,q\ell m}=\sum_\nu\sum_{p:\ell_p=\ell}\sum_n w^{(\nu)}_{qnp}\mathcal B^{(\nu)}_{i,npm}.
\end{equation}
The message that $j$ sends to $i$ leaves out the part that came from $i$ itself in first order, $I_{j\to i}=I_j-f_{0,j}w^{(1)}\phi_{ji}$, which is exact and costs one term per bond. 
The second layer couples the pair object with the message, $\psi_{ij}=\phi_{ij}\otimes I_{j\to i}$ (the first $n_{\mathrm{msg}}$ channels and degrees $\ell\le\ell_{\mathrm{feat}}$ of $\phi_{ij}$; channel $q$ of the message meets channel $q$ of the pair object), and sums over neighbours, $A^{(2)}_i=f_{1,i}\sum_j\psi_{ij}$. 
Its invariants are its scalars, the Gram matrices $P_\ell^\top P_\ell$ of projections $P_\ell=A^{(2)}_\ell W^{\mathrm G}_\ell$ to $r_{\mathrm{G}}$ columns, and the third-order invariants of projections $Q_\ell=A^{(2)}_\ell W^{\mathrm B}_\ell$ to $r_{\mathrm{B}}$ columns, a low-rank bispectrum~\cite{dpa-4c}.

The factors $f_{0,i}$ and $f_{1,i}$ of Eq.~\eqref{eq:gate} scale the two bond sums. 
The reference magnitude $\sigma_t$ of layer $t$ is determined layer by layer on the first training batches before training and kept fixed afterwards, also by fine-tunes. 
For the core switch, the atomic ZBL energy $E^{\mathrm{ZBL}}_i$ (below) enters
\begin{equation}
    S_i=1-10x^3+15x^4-6x^5,\ x=\min(E^{\mathrm{ZBL}}_i/z_{\mathrm c},1),
\end{equation}
with $z_{\mathrm c}=10$\,eV. $S_i$ and its first two derivatives are continuous, so energy, forces and their derivatives stay smooth.

The invariants $\mathbf x_i=\mathbf x^{(1)}_i\oplus\mathbf x^{(2)}_i$ are standardised with means and standard deviations of the training data and passed through one network shared by all elements,
\begin{equation}
    \begin{split}
        \mathbf h_i={}&\mathrm{SiLU}\bigl(\mathbf W_3\,\mathrm{SiLU}(\mathbf W_2\,\mathrm{SiLU}(\mathbf W_1\mathbf x'_i+\mathbf b_1+\boldsymbol\alpha_{Z_i})
        \\&+\mathbf b_2)+\mathbf b_3\bigr),
    \end{split}
\end{equation}
with one hidden layer in XS and $\mathbf x'_i$ the standardised invariants. 
The element of atom $i$ enters through the bias $\boldsymbol\alpha_{Z_i}$ and an output vector $\mathbf u_{Z_i}$. 
The energy is
\begin{equation}
    E=\sum_i\Bigl[E^0_{Z_i}+s\,S_i\,\mathbf h_i\cdot\mathbf u_{Z_i}+E^{\mathrm{ZBL}}_i\Bigr],
\end{equation}
with trainable reference energies $E^0_Z$, initialised by a least-squares fit to the composition, and a fixed energy scale $s$. 
The screened nuclear repulsion~\cite{zbl1985},
\begin{equation}
    E^{\mathrm{ZBL}}_i=\frac12\sum_j V_{\mathrm{ZBL}}(r_{ij})\,\tau(r_{ij}/r^{\mathrm s}_{ij})
\end{equation}
is switched off smoothly by 
\begin{equation}
    \tau(u)=\begin{cases}1-28u^6+48u^7-21u^8, & u<1,\\ 0, & u\ge 1,\end{cases}
\end{equation}
at $r^{\mathrm s}_{ij}=0.65\,(R_{\mathrm{cov},i}+R_{\mathrm{cov},j})$. 
Reference energies are added in double precision, because they are orders of magnitude larger than the learned part. 
Forces $\mathbf F_i=-\partial E/\partial\mathbf r_i$ and the stress $\boldsymbol\sigma=V^{-1}\partial E/\partial\boldsymbol\varepsilon$ follow by automatic differentiation, so the forces are conservative by construction. 
The models for molecules (OMOL, trained on \textsc{OMol25}~\cite{omol}) also take the total charge and the spin multiplicity of the structure as input: each is mapped to a learned vector, and both vectors are added to the element vector of every atom.
Supplementary Note~1 lists the hyperparameters of the six sizes. 
Supplementary Note~4 checks the implementation in double precision: energy, forces and stress are invariant under rotation, permutation and lattice translation to $10^{-11}$ or better, forces and stress agree with finite differences of the energy to $10^{-7}$ relative, and every internal tensor (the bond sums of both layers, the products and the messages) transforms under rotation as its degree requires, to $5\times10^{-15}$ relative in all six sizes.

\subsection{Training data and recipe}
\label{sec:m-training}

The universal materials models are trained on OAM, the union of \textsc{OMat24}~\cite{omat24}, its subsampled \textsc{Alexandria} counterpart \textsc{sAlex}~\cite{omat24,alexandria} and \textsc{MPtrj}~\cite{chgnet}: $110.6$~million training structures with $2.0\times10^9$~atoms of $89$~elements, $2$~\% of each source is held out as whole data files (shards of $2$-$5\times10^4$ structures, chosen with a fixed seed), so that all runs share one validation set.

The three sources differ in their DFT settings, and only \textsc{MPtrj} and \textsc{sAlex} share those of the \textsc{Materials Project}, so training runs in two stages. 
The first uses all sources; energies of the \textsc{OMat24} subsets receive a learned offset per element, so that they can teach forces and stresses without pulling the energy scale away from that of \textsc{MPtrj} and \textsc{sAlex}, by which the best checkpoint is selected. 
The second continues from this checkpoint at a tenth of the learning rate in one of two directions: on \textsc{MPtrj} and \textsc{sAlex} alone, after a least-squares refit of the reference energies (the MPA models, on the energy scale of the \textsc{Materials Project} and of \textsc{Matbench} Discovery), or on the \textsc{OMat24} subsets, with the learned offsets folded into the reference energies (the OMAT models, on \textsc{OMat24}'s scale). 
XS to L train $30$~epochs in the first stage and $15$ (MPA) or $5$ (OMAT) in the second; XL and XXL train $8$, $10$ (MPA) and $3$ (OMAT).

All models share one recipe. 
Matrix-shaped weights are updated with Muon~\cite{liu2025muon} (peak learning rate $2.5\times10^{-3}$, Nesterov momentum $0.95$, five Newton-Schulz iterations, update size matched to AdamW), all other parameters with AdamW~\cite{loshchilov2019} with AMSGrad~\cite{reddi2018} (peak $3\times10^{-3}$); weight decay is $10^{-3}$, gradients are clipped at a norm of $5$, and the learning rate rises linearly over $2{,}000$ steps and then follows a cosine to $2$~\% of its peak. 
The loss is the mean absolute error of the energy per atom, the force vectors and the stress, weighted $1$, $1$ and $3$ (forces $0.2$ in the MPA stage). 
Batches hold up to $8{,}192$~atoms or $512$~structures. 
Matrix products use the reduced-precision TF32 mode of the GPU, positions and strains float32, energy sums float64. 
The best checkpoint is chosen by the validation energy error of the raw weights. 
These settings came from a search on \textsc{MatPES} as a proxy (Supplementary Note~2).

The other datasets use the same recipe with their own lengths: \textsc{MatPES}~\cite{matpes} (PBE and r$^2$SCAN, the official split, $1{,}000$~epochs), \textsc{OMol25}~\cite{omol} ($1.0\times10^8$ structures of $83$~elements, $12$~epochs, with total charge and spin multiplicity as inputs) and \textsc{SolidAcid-0}~\cite{solidacid0} ($880{,}000$~frames of $55$~solid acids, $50$~epochs). 
OMOL models exist in sizes XS to XL and SOLIDACID models in S, M and L. 
Training ran on single \textsc{Nvidia} B200, H100 and H200 GPUs; the GPU-hours of every model are listed in Supplementary Note~2.

\subsection{Committees and weight averages}
\label{sec:m-ensembles}

We release two kinds of ensembles. 
The committee of the L size, in the MPA and OMAT versions, consists of five models that differ in the random initialisation and the batch order, each trained through all stages on its own; its spread estimates the error~\cite{zhang2020dpgen,schran2020}, at five times the cost. 
The weight average~\cite{wortsman2022soups,wortsman2022wise}, released for L (MPA and OMAT) and XXL (MPA), costs nothing at inference time: ten second-stage runs start from the same first-stage checkpoint, differ only in the order of their batches, and their weights are averaged with equal weights, following the weight averaging of fine-tuned models in image classification~\cite{wortsman2022soups,wortsman2022wise}.

\subsection{Fine-tuning}
\label{sec:m-ft}

Fine-tuning starts from any released model and offers four modes: all weights train (naive), a low-rank update of every weight matrix~\cite{hu2022lora} (rank~$8$, merged on saving), a second readout head trained next to one that replays the foundation on $1{,}000$~\textsc{MPtrj} structures (multi-head), and training the same architecture from random weights (scratch). 
Before training, the reference energies of the elements in the training set are corrected by a least-squares fit of the difference between reference and foundation energies, and they stay fixed afterwards. 
The loss is a weighted sum of squared errors, minimised by default with Adam and AMSGrad at a peak learning rate of $3\times10^{-3}$ (LoRA $10^{-2}$) for $200$~epochs. 
The comparison in Sec.~\ref{sec:res-ft} repeats the protocol of our earlier study~\cite{atk}, with errors on held-out trajectories.

\subsection{Inference engines}
\label{sec:m-engines}

On GPUs, \textsc{ILMlip} is evaluated by \textsc{JAX}~\cite{jax2018}: one molecular-dynamics step is one compiled call on a dense table of neighbours per atom, with bond sums, products and the bispectrum in the fused segmented-polynomial kernels of \textsc{cuEquivariance}~\cite{cuequivariance}. 
The neighbour list is a Verlet list with a skin of $1.5$~{\AA} (adjustable), built on the host with \textsc{vesin}~\cite{vesin,metatrain} in a background thread.
The kernels take the bonds of one atom in groups of four (two for the largest model) as one element of their batch, which leaves the result unchanged and makes the model call of S $13$~to $20$~\% faster at $10^4$ to $10^5$~atoms (Supplementary Note~5). 
Matrix products run in TF32; positions, lattice translations and strains are kept in float32, because TF32 rounding of the geometry produced size-dependent force errors.

On CPUs, a CPP engine parallelised with OpenMP uses the same Verlet list in float32 or float64. 
In \textsc{LAMMPS}~\cite{lammps}, \textsc{ILMlip} runs as an ML-IAP model on the KOKKOS GPU package~\cite{kokkos,lammps_kokkos}, and the messages of the second layer cross processor boundaries through \textsc{LAMMPS}'s ghost-atom exchange. 
In double precision, the \textsc{JAX} and CPP engines agree with the \textsc{PyTorch}~\cite{paszke2019,ansel2024} training model to better than $10^{-12}$~eV/{\AA}, and \textsc{LAMMPS} to $5\times10^{-12}$~eV/{\AA} on cells of $8$ to $576$~atoms (Supplementary Note~4). 
In production the network runs in single precision with TF32 matrix products on GPUs, which changes forces on average by $0.9$ (S) and $0.3$~meV/{\AA} (XL) against double precision, $0.8$ and $0.4$~\% of the force error of these models on held-out structures (Supplementary Note~4).

\subsection{Many simulations in one call}
\label{sec:m-batched}

\textsc{ILMlip} evaluates the structures of any number of independent simulations in one call, as one graph of disconnected parts, each with its own cell, strain and neighbour list, and the results equal those of separate calls. 
Path-integral simulations use this through \textsc{i-PI}~\cite{ipi3}, which hands all beads to the model at once, and plain \textsc{ASE} simulations through a driver that uses only \textsc{ASE}'s public calculator interface, so that any integrator, optimiser or elastic-band method works unchanged. 
The gains are given in Sec.~\ref{sec:res-speed} and Supplementary Note~5.

\subsection{Benchmarks}
\label{sec:m-bench}

Speed is measured as atom-steps per second of velocity-Verlet molecular dynamics of fcc copper in \textsc{ASE} (lattice constant $3.615$~{\AA}, $300$~K, $1$~fs), from $48$ to $10^8$~atoms on one \textsc{Nvidia} B200, with every model in the precision and engine its authors distribute.
The time of a model call alone is measured on $97{,}556$~atoms of the perfect lattice and, for the comparison with \textsc{MACE-MP-0} (small) and \textsc{MACE-OMAT-0} (medium, both float32 as distributed), on $10{,}976$~atoms, the largest cell all of them fit, with $20$~untimed and $100$ timed calls (Supplementary Note~5).
\textsc{Matbench} Discovery~\cite{matbench} is run with the shared runners of the benchmark repository on CPUs, with FIRE~\cite{bitzek2006} and a Fr{\'e}chet cell filter to a maximum force of $0.02$~eV/{\AA}.
Melting points come from solid-liquid coexistence: a two-phase cell of each crystal is simulated at constant pressure at $13$~temperatures between $0.6$ and $1.4$~times the experimental melting point, and the melting point is where the energy drift changes sign.
The experimental melting points and the results of other universal potentials are taken from the job archive of \textsc{MeltBench}~\cite{meltbench}, downloaded on 21~November 2026, whose runs use the small-cell coexistence method \textsc{SLUSCHI} instead.
Stability under heating follows \textsc{MLIPArena}~\cite{mliparena}, the amorphous test sets \textsc{AM26}~\cite{fragapane2026am26}, and the certification the default suite of \textsc{pescert}~\cite{haenseroth2026pescert} in double precision on CPUs.
The production simulations of the fine-tuned models use a Nos{\'e}-Hoover chain thermostat~\cite{martyna1992nhc} with a $0.5$~fs time step for up to $3$~ns: CDP at five temperatures between $510$ and $660$~K, KOH at $333$~K, Li$_{13}$Si$_4$ at $500$~K, L-pyroglutamic acid and phenol in water at $300$~K, and an elastic band of the S jump in MoS$_2$~\cite{henkelman2000neb}; diffusion coefficients are fitted to the mean squared displacement as in Ref.~\onlinecite{atk}. 
Details of all protocols are in the Supplementary Information.

\section*{Code Availability}

The \textsc{ILMlip} potential is available at \url{https://github.com/jhaens/ilmlip}.

The used third-party potentials 
\textsc{DeePMD}-kit~\cite{deepmd_v2, deepmd_v3, dpa, dpa3, dpa4, dpa-4c},
\textsc{GRACE}~\cite{grace_1,grace_2},
\textsc{MACE}~\cite{mace,macemp0,mace_2,mace_omat_mh},
\textsc{MatRIS}~\cite{matris},
\textsc{MatterSim}~\cite{mattersim}, 
\textsc{NEP89}~\cite{nep89},
\textsc{NequIP}~\cite{batzner2022,nequi_allegro_software},
\textsc{ORB}~\cite{orb,orbv3},
\textsc{SevenNet}~\cite{sevennet,sevennet_mf,sevennet_omni,sevennet-nano}
as well as the probed benchmarks 
\textsc{AM26}~\cite{fragapane2026am26},
\textsc{Matbench}~\cite{matbench},
\textsc{MeltBench}~\cite{meltbench},
\textsc{MLIPArena}~\cite{mliparena},
\textsc{pescert}~\cite{haenseroth2026pescert}
and the used simulation engines
\textsc{ASE}~\cite{ase},
\textsc{i-PI}~\cite{ipi3},
\textsc{LAMMPS}~\cite{lammps,lammps_kokkos},
\textsc{CP2K}~\cite{cp2k_1,cp2k_2,cp2k_3,cp2k_4,cp2k_5,cp2k_quickstep,cp2k_orb_trans}
and the libraries for the tensor operations and neighbour lists
\textsc{PyTorch}~\cite{paszke2019},
\textsc{cuEquivariance}~\cite{cuequivariance},
\textsc{JAX}~\cite{jax2018} and
\textsc{vesin}~\cite{vesin,metatrain}
are available from their respective repositories.
The used datasets
(subsampled) \textsc{Alexandria}~\cite{alexandria,omat24},
(subsampled) \textsc{Materials Project}~\cite{mp_1,mp_2,chgnet},
\textsc{MatPES}~\cite{matpes},
\textsc{OpenMaterials2024}~\cite{omat24},
\textsc{OpenMolecules2025}~\cite{omol},
\textsc{SolidAcid-0}~\cite{solidacid0}
are available from their respective databases.

\section*{Data Availability}

The \textsc{ILMlip} large-scale pretrained models are available at \url{https://github.com/jhaens/ilmlip-models}.
The exact version of the potential code and the corresponding models are available at \url{https://doi.org/10.5281/zenodo.24515567}.
The fine-tuning datasets are available from~\onlinecite{atk}.

\section*{Acknowledgments}

We thank the staff of the Compute Center of the Technische Universit{\"a}t Ilmenau, especially Mr.~Henning~Schwanbeck for providing an excellent research environment.
J.~H{\"a}nseroth thanks M.~Otto and P.~Merle for testing the large-scale pretrained models and M.~Gro{\ss}mann, M.~Grunert, E.~Pabst and J.~L.~Wolf for their valuable feedback on the architecture of \textsc{ILMlip}.
This work is supported by the doctoral scholarship of the German Academic Scholarship Foundation.

\section*{Competing interests}

The authors declare no competing interests.

\section*{Author contributions}

J.H.~conceived the study, developed the implementation of the potential with assistance from LLM-based coding tools, trained the universal models, and performed the benchmark calculations. 
Subsequent software development was supported by Claude~Code, powered by Claude~Opus~5.5 (extra-high and maximum reasoning effort), which assisted with test design and implementation, code refinement, debugging, documentation design and implementation, software testing, translation from PyTorch to CPP, and performance optimization. 
The development process remained under the direction of J.H., who reviewed and validated the resulting implementations.
J.H.~analysed the data.
J.H.~produced all figures and wrote the first draft of the manuscript.
C.D.~supervised the work.
All authors revised and approved the manuscript.

\bibliography{references}

\end{document}


\title{Supplementary Information for ``ILMlip: fast and universal machine-learning interatomic potentials''}

\author{Jonas H{\"a}nseroth}
\email{jonas.haenseroth@tu-ilmenau.de}
\affiliation{Theoretical Solid State Physics, Institute of Physics, Technische Universit{\"a}t Ilmenau, 98693 Ilmenau, Germany}

\author{Christian Dre{\ss}ler}
\affiliation{Theoretical Solid State Physics, Institute of Physics, Technische Universit{\"a}t Ilmenau, 98693 Ilmenau, Germany}

\date{\today}

\maketitle

All abbreviations are defined in the main text; abbreviations used only here are defined at first use.

\tableofcontents
\newpage
\clearpage

\newpage
\section*{Supplementary Note 1: Model sizes}

\begin{table}[h]
    \caption{\textbf{Hyperparameters of the \textsc{ILMlip} model sizes.} 
    $n_\ell$: channels of degree $\ell$ in the first layer. 
    Bounds in all sizes: $\kappa=16$, $a_0=0.05$, core switch $z_{\mathrm c}=10$~eV, feature squash off; ZBL switched off at $0.65$~$(R_{\mathrm{cov},i}+R_{\mathrm{cov},j})$.}
    \label{tab:si-sizes}
    \centering
    \small
    \begin{tabular}{lrrrrrr}
        \toprule
         & XS & S & M & L & XL & XXL \\
        \midrule
        cut-off & \multicolumn{4}{c}{4-6\,\AA{} per element} & 6\,\AA & 6\,\AA \\
        $\ell_{\max}/\ell_{\mathrm{feat}}/\ell_{\mathrm{msg}}$ & 2/1/1 & 2/1/1 & 2/1/1 & 2/2/1 & 3/2/2 & 3/3/3 \\
        $n_{\mathrm{base}}$ & 8 & 8 & 8 & 6 & 16 & 16 \\
        taper $n_0/n_1/n_2(/n_3)$ & 24/12/6 & 32/16/8 & 32/16/8 & 32/16/16 & 96/48/48/12 & 128/64/64/64 \\
        body order (factors $\nu$) & 4 (3) & 4 (3) & 4 (3) & 4 (3) & 4 (3) & 3 (2) \\
        $n_{\mathrm{msg}}$ & 12 & 14 & 14 & 15 & 48 & 64 \\
        $r_{\mathrm{G}}/r_{\mathrm{B}}$ & 12/6 & 14/8 & 14/8 & 15/8 & 48/16 & 64/16 \\
        readout & $1\times256$ & $3\times256$ & $3\times512$ & $3\times512$ & $3\times1{,}024$ & $3\times2{,}048$ \\
        element vector $d_z$ & 32 & 32 & 32 & 32 & 64 & 64 \\
        invariants $n_{\mathrm{feat}}$ & 306 & 406 & 406 & 622 & 4\,376 & 9\,152 \\
        parameters & $140{,}456$ & $300{,}612$ & $849{,}732$ & $963{,}587$ & $6{,}953{,}980$ & $27{,}822{,}308$ \\
        \quad in the readout & $56$~\% & $78$~\% & $86$~\% & $88$~\% & $95$~\% & $97.5$~\% \\
        \bottomrule
    \end{tabular}
\end{table}

\clearpage
\section*{Supplementary Note 2: Training}

\begin{table}[htbp]
    \caption{\textbf{Training cost and held-out errors of every released model.} 
    Mean absolute errors: energy in meV/atom, force in meV/{\AA}; held-out sets in Supplementary Note~6. 
    GPU-hours: wall-clock hours of the training run.
    Models that trained together on one GPU share its wall time, so one of them alone would cost less: OAM XS, M and L (one B200; $45.7$~h for the three first stages together, not the sum of the entries), \textsc{MatPES} XS to L (one GPU per functional), \textsc{OMol25} XS with L and S with M, \textsc{SolidAcid-0} S, M and L. 
    OAM S: first stage on a \textsc{Nvidia} H100 NVL and then an H200 NVL, second stage on an H200 NVL. 
    Last OAM row: the S model submitted to \textsc{Matbench} Discovery.}
    \label{tab:si-training}
    \centering\small
    \setlength{\tabcolsep}{5pt}
    \begin{tabular}{lcrrrlcc}
        \toprule
        \multicolumn{8}{l}{OAM (first stage (1), then MPA or OMAT)} \\
         & epochs & \multicolumn{3}{c}{GPU-hours} & GPU & \multicolumn{2}{c}{$E$ / $F$} \\
         & 1 / MPA / OMAT & 1 & MPA & OMAT & & MPA & OMAT \\
        \midrule
        XS & 30 / 15 / 5 & 41.2 & 2.0 & 5.2 & B200, shared & 28.7 / 73.9 & 26.9 / 123.2 \\
        S & 30 / 15 / 5 & 22.8 & 1.1 & 2.8 & H100 / H200 NVL & 22.3 / 63.1 & 22.8 / 112.9 \\
        M & 30 / 15 / 5 & 44.6 & 2.2 & 5.6 & B200, shared & 19.5 / 59.2 & 20.8 / 108.1 \\
        L & 30 / 15 / 5 & 45.7 & 2.3 & 5.7 & B200, shared & 18.0 / 54.9 & 19.3 / 101.1 \\
        XL & 8 / 10 / 3 & 15.9 & 2.2 & 4.5 & B200 & 11.8 / 39.2 & 13.1 / 76.3 \\
        XXL & 8 / 10 / 3 & 31.0 & 4.2 & 8.7 & B200 & 10.7 / 40.1 & 11.6 / 67.3 \\
        S, \textsc{Matbench} & 8 / 10 / -- & 6.2 & 0.7 & -- & H100 NVL, B200 & 23.9 / 66.5 & -- \\
        \midrule
        \multicolumn{8}{l}{\textsc{MatPES} ($1{,}000$~epochs)} \\
         & & \multicolumn{2}{c}{PBE} & \multicolumn{2}{c}{r$^2$SCAN} & \multicolumn{2}{c}{$E$ / $F$} \\
         &  & \multicolumn{2}{c}{GPU-hours} & \multicolumn{2}{c}{GPU-hours} & PBE & r$^2$SCAN \\
        \midrule
        XS &  & \multicolumn{2}{c}{6.4} & \multicolumn{2}{c}{5.4} & 39.2 / 134.7 & 38.6 / 165.0 \\
        S &  & \multicolumn{2}{c}{6.8} & \multicolumn{2}{c}{5.8} & 31.9 / 125.1 & 31.1 / 156.1 \\
        M &  & \multicolumn{2}{c}{6.9} & \multicolumn{2}{c}{5.9} & 29.2 / 130.5 & 28.1 / 165.5 \\
        L &  & \multicolumn{2}{c}{6.9} & \multicolumn{2}{c}{5.9} & 27.3 / 124.5 & 26.6 / 159.7 \\
        XL &  & \multicolumn{2}{c}{8.6} & \multicolumn{2}{c}{6.7} & 21.1 / 112.5 & 19.3 / 144.6 \\
        XXL &  & \multicolumn{2}{c}{17.1} & \multicolumn{2}{c}{14.3} & 19.6 / 97.7 & 17.6 / 126.7 \\
        \midrule
        \multicolumn{8}{l}{\textsc{OMol25} ($12$~epochs) and \textsc{SolidAcid-0} ($50$~epochs)} \\
         & & \multicolumn{2}{c}{\textsc{OMol25}} & \multicolumn{2}{c}{\textsc{SolidAcid-0}} & \multicolumn{2}{c}{$E$ / $F$} \\
         & & \multicolumn{2}{c}{GPU-hours} & \multicolumn{2}{c}{GPU-hours} & \textsc{OMol25} & \textsc{SolidAcid-0} \\
        \midrule
        XS & & \multicolumn{2}{c}{33.1} & \multicolumn{2}{c}{--} & 23.6 / 70.2 & -- \\
        S & & \multicolumn{2}{c}{29.0} & \multicolumn{2}{c}{67.1} & 15.1 / 58.5 & 0.84 / 20.5 \\
        M & & \multicolumn{2}{c}{29.6} & \multicolumn{2}{c}{63.0} & 12.7 / 51.0 & 0.72 / 18.7 \\
        L & & \multicolumn{2}{c}{37.0} & \multicolumn{2}{c}{67.4} & 11.2 / 45.2 & 0.62 / 16.3 \\
        XL & & \multicolumn{2}{c}{75.4} & \multicolumn{2}{c}{--} & 4.55 / 22.7 & -- \\
        \bottomrule
    \end{tabular}
\end{table}

The four other members of the L committee trained in packs of four: $75.1$~h for the first stage and $3.6$~h (MPA) or $9.4$~h (OMAT) for the second, as wall time of the shared GPU (\textsc{Nvidia} B200). 

On \textsc{MatPES} the validation force error of XL and XXL passes a minimum after about half of the steps and rises afterwards while the training loss and the energy error keep falling drastically.
\textsc{OMol25} XL has single evaluations with several times the running force error (up to about $300$ against $40$~meV/{\AA}), which recover at the next one. 
The validation energy error of \textsc{SolidAcid-0} fluctuates by an order of magnitude.

Batches of $8{,}192$~atoms or $512$~structures gave the error of batches of $2{,}048$~atoms or $256$~structures and ran $25$ to $30$~\% faster. 
The released \textsc{MatPES} models used a weight decay of $10^{-2}$ instead of $10^{-3}$, which lowers the energy error by $1$ to $3$~meV/atom for XS to L, leaves XL and XXL equal and raises the XL force error by $7$~meV/\AA; otherwise they differ from the recipe runs of Supplementary Table~\ref{tab:si-training} by at most $3$~meV/atom and $7$~meV/{\AA}.

\begin{figure}[tbp]
    \centering
    \includegraphics{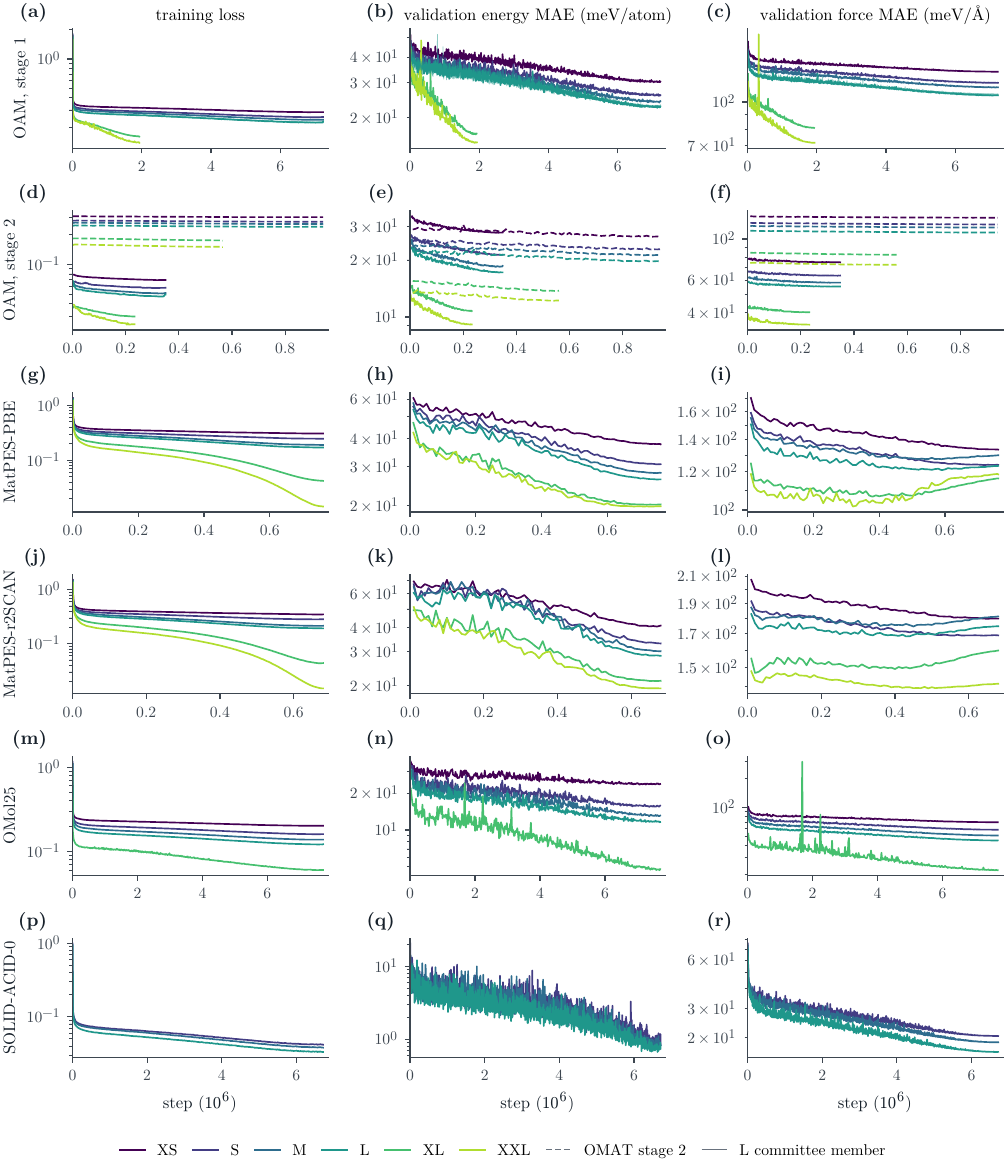}
    \caption{\textbf{Pretraining performance.} 
    Training of every model family over the optimiser steps: training loss (weighted MAE of the batches, rolling median), validation energy and force MAE; rows: OAM stage~1, OAM stage~2 (MPA solid, OMAT dashed), \textsc{MatPES}-PBE, \textsc{MatPES}-r$^2$SCAN, \textsc{OMol25}, \textsc{SolidAcid-0}.}
\label{fig:si-training}
\end{figure}

\clearpage
\section*{Supplementary Note 3: Ablations of the architecture}
\label{si:ablation}

Ablations were run on \textsc{MatPES}-PBE for $200$~epochs around the size class of L.
Validation errors are given as energy and force MAE; three seeds of the baseline scatter by $0.8$~meV/atom and $2.2$~meV/{\AA}.

\begin{table}[h]
    \caption{\textbf{Architecture ablations.}}
    \label{tab:si-abl}
    \centering\small\setlength{\tabcolsep}{5pt}
    \begin{tabular}{p{4.0cm}p{6.4cm}p{5.6cm}}
        \toprule
        change & $E$ (meV/atom) / $F$ (meV/{\AA}) & reading \\
        \midrule
        baseline (3 seeds) & 28.5 / 122.7 & (with random element embedding) \\
        pair FiLM off & 30.6 / 131.3 & FiLM pays \\
        periodic element code & 27.4 / 118.7 & gain, adopted \\
        uniform cut-off & & \\ 
        \hspace{2cm} $4$~{\AA}  & 39.6 / 131.8 & \\
        \hspace{2cm} $5$~{\AA}  & 28.3 / 123.8 & \\
        \hspace{2cm} $6$~{\AA}  & 26.9 / 125.8 &  element cut-offs \\
        taper off & 28.0 / 121.7 & no accuracy change, $1.3\times$ the invariants \\
        second layer from its own pair object & 28.4 / 121.9 & sharing costs nothing \\
        back-flow kept & 29.5 / 126.6 & no measurable effect \\
        full product table instead of Gram + bispectrum & 29.6 / 126.6 & no gain at $4\times$ the invariants \\
        no message layer & 31.1 / 135.8 & the message layer pays \\
        three layers (order 2) & 28.4 / 121.3 & no gain at $4.3\times$ the time \\
        readout depth & & \\ 
        \hspace{2cm} 1 & 34.8 / 133.5 & \\
        \hspace{2cm} 2 & 30.7 / 125.5 & \\
        \hspace{2cm} 4 & 28.2 / 125.5 & three layers \\
        readout width & & \\
        \hspace{2cm} 128  & 35.9 / 132.3 & \\
        \hspace{2cm} 256  & 32.0 / 126.1 & \\
        \hspace{2cm} $1{,}024$ & 27.0 / 127.2 & \\
        \hspace{2cm} $2{,}048$ & 26.1 / 130.6 & \\
        ZBL off & 29.3 / 125.2 & neutral on \textsc{MatPES} \\
        $\ell_{\max}$ & & \\
        \hspace{2cm} 1 & 34.4 / 150.7 & \\
        \hspace{2cm} 3 & 26.8 / 118.5 & \\
        \hspace{2cm} 4 & 26.3 / 117.9 & $2$ or $3$ compete with the increased cost \\
        \bottomrule
    \end{tabular}
\end{table}

The two bounds were not part of this series and were ablated separately on S and L with the recipe of the main text, with both, with the gate only, with the core switch only and with neither (Supplementary Table~\ref{tab:si-bounds}; test split of $21{,}660$~structures, one run per cell).
Two seeds of the production recipe differ by $0.43$ and $0.58$~meV/atom and by $1.1$ and $0.7$~meV/{\AA} for S and L.
The four variants lie within $0.6$ (S) and $0.2$~meV/atom (L) of each other in energy and within $1.4$ and $1.5$~meV/{\AA} in force.
Of the other settings, gate scales $\kappa=8$ and $32$ are within $0.2$~meV/atom and $1.6$~meV/{\AA} of $\kappa=16$, a gate on the norm of the bond sum instead of its root mean square, at $\kappa=8$, is much worse (S $50.1$~meV/atom and $171$~meV/{\AA}, L $32.5$~meV/atom and $132$~meV/{\AA}).

\begin{table}[htbp]
    \caption{\textbf{Ablation of the gate and the core switch.}
    On \textsc{MatPES}-PBE, $200$~epochs, test errors ($E$ in meV/atom, $F$ in meV/{\AA}, $S$ in meV/{\AA}$^3$).
    Production: both bounds.
    One run per cell.}
    \label{tab:si-bounds}
    \centering\small\setlength{\tabcolsep}{5pt}
    \begin{tabular}{llrrr}
        \toprule
        size & variant & $E$ & $F$ & $S$ \\
        \midrule
        S & production (gate, core switch) & 35.11 & 127.7 & 5.74 \\
        & production, seed 1 & 35.55 & 128.8 & 5.79 \\
        & core switch only & 35.34 & 128.4 & 5.86 \\
        & gate only & 34.70 & 127.0 & 5.73 \\
        & neither & 35.08 & 126.9 & 5.77 \\
        \midrule
        L & production (gate, core switch) & 28.71 & 121.2 & 5.24 \\
        & production, seed 1 & 29.29 & 121.9 & 5.26 \\
        & core switch only & 28.75 & 121.2 & 5.27 \\
        & gate only & 28.87 & 119.7 & 5.21 \\
        & neither & 28.68 & 120.5 & 5.24 \\
        \bottomrule
    \end{tabular}
\end{table}

\clearpage
\section*{Supplementary Note 4: Properties of the bounds and checks of the implementation}
\label{si:proofs}

The scripts are shipped with the code (the \texttt{checks} folder of the \textsc{ILMlip} repository); \texttt{bounds\_sympy.py} proves the statements symbolically and compares them with the package's functions, \texttt{invariance\_check.py} tests a trained S model numerically in double precision.

For the gate, with $x=\sqrt{\langle A^2\rangle}/(\sigma\kappa)$, the root-mean-square component of $f A$ is $a_0\kappa\,x/\sqrt{1+x^2}$: strictly increasing, below $a_0\kappa$ for every density and tending to it; $f\to a_0/\sigma$ for $\langle A^2\rangle\to0$.
At the reference density ($\rho=1$) the factor is $(1+\kappa^{-2})^{-1/2}=0.99805$ of its low-density value for $\kappa=16$.

\begin{figure}[h]
    \centering
    \includegraphics[width=\linewidth]{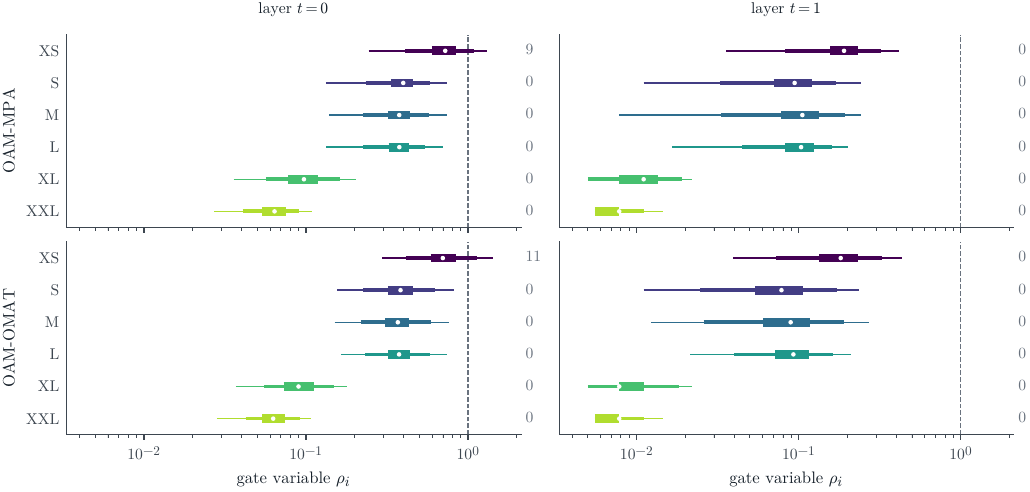}
    \caption{\textbf{Distribution of the gate variable.}
    The variable $\rho_i$ (bond sum in units of its value in the freshly initialised network) of the OAM-MPA and OAM-OMAT models, XS to XXL, for the first ($t=0$) and the second layer ($t=1$), on $2{,}000$ held-out structures of each stage.
    Bars: $1$-$99$~\%, $5$-$95$~\% and $25$-$75$~\% of the atoms, dot: median; grey number: percentage of atoms with $\rho_i>1$; dashed line: $\rho_i=1$.
    Values below $0.005$ are at the resolution of the single-precision gate.}
    \label{fig:si-gate-rho}
\end{figure}

Supplementary Figure~\ref{fig:si-gate-rho} shows where the released models sit on this curve: the distribution of $\rho_i$ of Eq.~(1) of the main text, the root-mean-square bond sum of atom $i$ in units of its value in the freshly initialised network, in both layers of the twelve OAM models on $2{,}000$ held-out structures of each stage.
In the first layer the median is $0.70$ and $0.72$ in XS, $0.37$ to $0.40$ in S to L and $0.06$ to $0.10$ in XL and XXL.
In the second layer the median is $0.18$ and $0.19$ in XS, $0.08$ to $0.11$ in S, M and L and at most $0.02$ in XL and XXL, where most atoms lie at the resolution of the single-precision gate.
A bond sum above the calibration value ($\rho_i>1$) occurs for $9$ to $11$~\% of the first-layer atoms of XS, for at most $0.1$~\% in S and M and nowhere else; no atom has $\rho_i>2$.
The bond sums of the trained models are thus smaller than those of the initial network, more so for the larger models; the gate scales the median atom by less than $0.1$~\% (at $\rho_i=\kappa=16$ by $29$~\%) and acts as a bound only for structures far from the training data.

The core switch $S(x)=1-10x^3+15x^4-6x^5$ has $S(0)=1$, $S(1)=0$, vanishing first and second derivatives at both ends and $S'=-30x^2(1-x)^2\le0$; the clipped switch is $C^2$.
The taper of the screened nuclear repulsion, $\tau(u)=(1-u)^3(1+3u+6u^2+10u^3+15u^4+21u^5)=1-28u^6+48u^7-21u^8$, is $C^2$ at $u=1$ and keeps the repulsion unchanged to $O(u^6)$ near the nuclei.

Invariance and conservative forces were tested numerically on a random $40$-atom triclinic cell of five elements and on a compressed one in which $28$ of $40$~atoms are inside the core switch (Supplementary Table~\ref{tab:si-checks}).

\begin{table}[h]
    \caption{\textbf{Numerical checks of a trained S model.}
    Double precision (float64), \textsc{PyTorch}; units $E$ eV, $F$ eV/{\AA}, $\sigma$ eV/{\AA}$^3$.}
    \label{tab:si-checks}
    \centering\small\setlength{\tabcolsep}{5pt}
    \begin{tabular}{lll}
        \toprule
        test & largest deviation & compressed cell \\
        \midrule
        rotation of positions and cell ($E$/$F$/$\sigma$) & $6\times10^{-14}$ / $3\times10^{-14}$ / $2\times10^{-15}$ & $\le10^{-11}$ \\
        permutation of atoms & $0$ / $7\times10^{-14}$ / $3\times10^{-16}$ & $\le10^{-11}$ \\
        lattice translation & $2\times10^{-13}$ / $2\times10^{-13}$ / $4\times10^{-16}$ & $9\times10^{-12}$ \\
        forces against finite differences (relative) & $3.8\times10^{-9}$ & $9.0\times10^{-8}$ \\
        stress against finite differences (relative) & $9.3\times10^{-9}$ & $2.5\times10^{-8}$ \\
        \bottomrule
    \end{tabular}
\end{table}

Invariant energies and equivariant forces do not by themselves show that every intermediate tensor transforms correctly, since an error in a coupling coefficient could cancel in the invariants.
\texttt{equivariance\_check.py} therefore passes a random structure and a rotated copy through the network and requires every degree-$\ell$ block of the gated bond sums of both layers, of every product function of the first layer (all orders and degrees) and of the messages to satisfy $X(R)=X\,D^{\ell}(R)$.
The Wigner matrices are obtained by least squares from the model's own real spherical harmonics, so no convention is assumed.
For all six sizes, including degrees up to three in XL and XXL, the largest relative deviation is $5\times10^{-15}$.

The engines (\textsc{JAX}, CPP, \textsc{LAMMPS}) agree with the \textsc{PyTorch} model to $10^{-11}$~eV/{\AA} or better in double precision (Supplementary Table~\ref{tab:si-engines}).
The test set holds $23$~cells, $20$ strained and rattled bulk crystals of $32$ to $96$~atoms, fcc Cu with $500$~atoms, NaCl with $512$ and water with $192$, evaluated by OAM-S-OMAT and OAM-XL-OMAT against the \textsc{JAX} engine in double precision.
In production, the TF32 matrix products change the forces by $0.9$ (S) and $0.3$~meV/{\AA} (XL) on average, and the TF32 export of \textsc{LAMMPS} deviates by the same order.

\begin{table}[htbp]
    \caption{\textbf{Agreement of the engines and precisions.}
    \textsc{JAX} and \textsc{PyTorch} against the \textsc{JAX} engine in double precision and \textsc{LAMMPS} against the \textsc{PyTorch} model in double precision, largest value of the three cells, mean column: largest cell mean).
    Forces in meV/{\AA}, energies in meV/atom, stress in MPa.}
    \label{tab:si-engines}
    \centering\small\setlength{\tabcolsep}{5pt}
    \begin{tabular}{llrrrr}
        \toprule
        model & engine, precision & max $|\Delta F|$ & mean $|\Delta F|$ & max $|\Delta E|$ & max $|\Delta\sigma|$ \\
        \midrule
        S & \textsc{JAX}, TF32 (production) & 15.5 & 0.90 & 2.47 & 146 \\
        & \textsc{JAX}, float32 & 0.045 & 0.0026 & 0.0029 & 0.07 \\
        & \textsc{PyTorch}, float64 (GPU) & $4\times10^{-11}$ & $4\times10^{-12}$ & $6\times10^{-12}$ & $2\times10^{-10}$ \\
        & CPP, float64 (CPU) & $5\times10^{-11}$ & $4\times10^{-12}$ & $3\times10^{-12}$ & $9\times10^{-11}$ \\
        & \textsc{LAMMPS}, float64 export & $5\times10^{-9}$ & $1\times10^{-9}$ & $9\times10^{-11}$ & $5\times10^{-8}$ \\
        & \textsc{LAMMPS}, TF32 export & 4.9 & 0.89 & 0.90 & 12 \\
        \midrule
        XL & \textsc{JAX}, TF32 (production) & 6.0 & 0.29 & 0.61 & 38 \\
        & \textsc{JAX}, float32 & 0.041 & 0.0023 & 0.0011 & 0.08 \\
        & \textsc{PyTorch}, float64 (GPU) & $6\times10^{-11}$ & $4\times10^{-12}$ & $5\times10^{-12}$ & $5\times10^{-10}$ \\
        & CPP, float64 (CPU) & $5\times10^{-11}$ & $3\times10^{-12}$ & $3\times10^{-12}$ & $2\times10^{-10}$ \\
        & \textsc{LAMMPS}, float64 export & $5\times10^{-9}$ & $1\times10^{-9}$ & $6\times10^{-11}$ & $4\times10^{-8}$ \\
        & \textsc{LAMMPS}, TF32 export & 1.6 & 0.29 & 0.22 & 5.9 \\
        \bottomrule
    \end{tabular}
\end{table}

\clearpage
\section*{Supplementary Note 5: Speed}
\label{si:speed}

Speed is reported as atom-steps per second: the number of atoms times the number of MD steps, divided by the wall time of the whole step, the model call and the rest of the step together.
The protocol is that of the Methods (fcc copper, $3.615$~{\AA}, start temperature $300$~K, velocity-Verlet with $1$~fs through \textsc{ASE}, $48$ to $10^7$ atoms, one \textsc{Nvidia} B200 with $183$~GB).
Every run has $20$ untimed steps and three repeats of $100$ timed steps, fewer at the largest sizes.
\textsc{ILMlip} runs with the trained \textsc{MatPES}-PBE weights of each size, in \textsc{JAX} with TF32 matrix products, fused kernels and \textsc{CUDA} graphs.
Every other model runs in the engine and precision of its distribution: \textsc{MACE} small (MP-0) and \textsc{MACE-OMAT} medium~\cite{macemp0} in single precision with \textsc{PyTorch}, \textsc{GRACE-2L-OAM}~\cite{grace_2} in double precision with \textsc{TensorFlow}, \textsc{SevenNet-0}~\cite{sevennet} and \textsc{DPA-4.1-Air} and \textsc{DPA-4.1C-Air}~\cite{dpa4,dpa-4c} through their own packages, and \textsc{NEP89}~\cite{nep89} through an \textsc{ASE} adapter of its \textsc{CUDA} code.

\subsection*{Scaling and the largest system on one GPU}
\label{si:speed-scaling}

Supplementary Table~\ref{tab:si-scaling} gives the throughput at $10^4$ and $10^5$ atoms, the peak over the twelve sizes run, and the largest size that fitted.
The main text compares S with \textsc{NEP89} and \textsc{DPA-4.1C-Air} at $10^4$ atoms; at the peak S is $4.7$ and $2.8$~times faster than the two, and XS is $5.6$ and $3.4$~times faster at $10^4$ atoms.
XL and XXL pay for their accuracy (Supplementary Note~6) in speed: at $10^4$ atoms they reach $0.59$ and $0.23$~times \textsc{NEP89} and $0.35$ and $0.14$~times \textsc{DPA-4.1C-Air}.
\begin{table}[htbp]
    \caption{\textbf{MD throughput through \textsc{ASE}.}
    In $10^6$ atom-steps/s (fcc Cu, one B200).
    Peak: maximum over the sizes that were run, with the size in brackets. 
    Not run (--). 
    ns/day: simulated time per day at $10^4$ atoms with $1$~fs per step.}
    \label{tab:si-scaling}
    \centering\small\setlength{\tabcolsep}{5pt}
    \begin{tabular}{lrrrrr}
        \toprule
        & $10^4$ atoms & $10^5$ atoms & peak (atoms) & ns/day & largest size run \\
        \midrule
        \textsc{ILMlip} XS & 12.5 & 20.4 & \textbf{20.4} ($10^5$) & 108 & $10^8$ \\
        \textsc{ILMlip} S & 11.6 & 18.3 & 18.3 ($10^5$) & 101 & $10^8$ \\
        \textsc{ILMlip} M & 10.6 & 16.4 & 16.4 ($10^5$) & 91.5 & $10^8$ \\
        \textsc{ILMlip} L & 8.02 & 11.6 & 11.6 ($10^5$) & 69.3 & $7\times10^7$ \\
        \textsc{ILMlip} XL & 1.31 & 1.64 & 1.64 ($10^5$) & 11.3 & $2\times10^7$ \\
        \textsc{ILMlip} XXL & 0.507 & 0.515 & 0.520 ($5\times10^4$) & 4.4 & $2\times10^7$ \\
        \midrule
        \textsc{NEP89} & 2.21 & 3.87 & 3.87 ($10^5$) & 19.1 & $2\times10^7$ \\
        \textsc{DPA-4.1C-Air} & 3.72 & 6.38 & 6.48 ($5\times10^5$) & 32.1 & $5\times10^6$ \\
        \textsc{GRACE-2L-OAM} & 0.0235 & 0.0251 & 0.0251 ($5\times10^4$) & 0.20 & $10^5$ \\
        \textsc{MACE} small & 0.0200 & -- & 0.0200 ($5\times10^3$) & 0.17 & $5\times10^4$ \\
        \textsc{MACE-OMAT} medium & 0.0126 & -- & 0.0127 ($5\times10^3$) & 0.11 & $10^4$ \\
        \textsc{SevenNet-0} & 0.0226 & -- & 0.0226 ($10^4$) & 0.19 & $10^4$ \\
        \textsc{DPA-4.1-Air} & 0.0322 & -- & 0.0322 ($10^4$) & 0.28 & $10^4$ \\
        \bottomrule
    \end{tabular}
\end{table}

Below $10^5$ atoms the throughput of \textsc{ILMlip} rises with the number of atoms because the fixed cost of a step (list check, driver) is spread over more atoms.
The V100 throughputs that Ref.~\onlinecite{dpa-4c} reports for \textsc{NEP89} in \textsc{GPUMD}~\cite{gpumd} and for \textsc{DPA-4.1C-Air} in its own engine ($7.0$ and $3.1$~M atom-steps/s for fcc Cu) cannot be compared with this table; the ratios above hold in \textsc{ASE}.

Supplementary Table~\ref{tab:si-limits} lists the largest fcc Cu system that each model ran on one B200, with ten timed steps per size.
The memory column gives what the card reports for the process, including \textsc{JAX}'s memory pool, and what \textsc{JAX} has allocated.
\begin{table}[h]
    \caption{\textbf{Largest system on one B200 (fcc Cu).}
    GPU memory at $10^6$ atoms: used by the process / allocated by \textsc{JAX} (GB); Not measured (--).}
    \label{tab:si-limits}
    \centering\small\setlength{\tabcolsep}{5pt}
    \begin{tabular}{lrrr}
        \toprule
        & largest size run & $10^6$ atom-steps/s there & memory at $10^6$ atoms \\
        \midrule
        \textsc{ILMlip} XS & $10^8$ & 14.0 & 69.8 / 27.5 \\
        \textsc{ILMlip} S & $10^8$ & 12.9 & 54.9 / 30.9 \\
        \textsc{ILMlip} M & $10^8$ & 11.8 & 72.1 / 37.1 \\
        \textsc{ILMlip} L & $7\times10^7$ & 8.99 & 72.1 / 44.0 \\
        \textsc{ILMlip} XL & $2\times10^7$& 1.26 & 72.8 / 28.6 \\
        \textsc{ILMlip} XXL & $2\times10^7$ & 0.489 & 76.1 / 27.6 \\
        \textsc{NEP89} & $2\times10^7$ & 3.44 & 7.05 / -- \\
        \textsc{DPA-4.1C-Air} & $5\times10^6$ & 6.42 & 26.0 / 9.1 \\
        \bottomrule
    \end{tabular}
\end{table}

\subsection*{Model call timing}
\label{si:speed-call}

One MD step costs the model call, the compiled network evaluated on the GPU together with the transfer of positions and forces between host and GPU, plus the rest of the step, the integrator and the driver (here \textsc{ASE}).
Supplementary Table~\ref{tab:si-call} gives both per atom: the model call timed alone on $97{,}556$~atoms of the perfect lattice (median of $100$ calls after $20$ untimed ones, stress off), and the MD step as the inverse throughput of velocity-Verlet MD at $10^5$ atoms (Supplementary Table~\ref{tab:si-scaling}).
The rest of the step costs about $0.014$~$\mu$s per atom for every model, $28$~\% of the step for XS and $6$~\% for XXL.
XL costs $14$~times and XXL $46$~times the model call of S with $23$ and $93$~times its parameters, because their $6$~{\AA} cut-off raises the bonds per atom of copper from $18$ to $78$ and they carry more channels and higher angular degrees (Sec.~\ref{si:speed-hparam}).

\begin{table}[htbp]
    \caption{\textbf{Cost per atom of one model call and of one MD step.}
    Model call on $97{,}556$~atoms of fcc Cu, MD step with velocity-Verlet through \textsc{ASE} at $10^5$ atoms, one B200.
    Share: the model call as a fraction of the MD step.}
    \label{tab:si-call}
    \centering\small\setlength{\tabcolsep}{5pt}
    \begin{tabular}{lrrrrrr}
        \toprule
        & \multicolumn{2}{c}{model call ($\mu$s/atom)} & MD step & share of & peak memory & bonds \\
        & GPU only & with transfers & ($\mu$s/atom) & model call & (GB) & per atom \\
        \midrule
        XS & 0.0342 & 0.0353 & 0.0490 & $72$~\% & 1.93 & 18 \\
        S & 0.0392 & 0.0407 & 0.0548 & $74$~\% & 2.12 & 18 \\
        M & 0.0456 & 0.0468 & 0.0610 & $77$~\% & 2.72 & 18 \\
        L & 0.0694 & 0.0706 & 0.0864 & $82$~\% & 3.02 & 18 \\
        XL & 0.562 & 0.565 & 0.610 & $93$~\% & 25.0 & 78 \\
        XXL & 1.816 & 1.824 & 1.942 & $94$~\% & 24.4 & 78 \\
        \bottomrule
    \end{tabular}
\end{table}

\begin{table}[htbp]
    \caption{\textbf{Cost per atom of one model call against \textsc{MACE}.}
    Energy and forces of $10{,}976$~atoms of fcc Cu, the largest cell that every model fits, on one B200; median of $100$ calls after $20$ untimed ones.
    GPU only: the compiled call on positions already on the GPU; with transfers: the call as a MD step makes it, for \textsc{MACE} including its neighbour list on the host.
    \textsc{ILMlip}: the \textsc{MatPES}-PBE weights in TF32 with fused kernels; \textsc{MACE-MP-0} (small) and \textsc{MACE-OMAT-0} (medium; mace-torch 0.3.13) in float32 without TF32 and cuEquivariance, as in the throughput benchmark.
    At $97{,}556$~atoms both \textsc{MACE} models exceed the $183$~GB of the GPU.}
    \label{tab:si-call-mace}
    \centering\small\setlength{\tabcolsep}{5pt}
    \begin{tabular}{lrrrrr}
        \toprule
        & & \multicolumn{2}{c}{model call ($\mu$s/atom)} & peak memory & bonds \\
        & parameters & GPU only & with transfers & (GB) & per atom \\
        \midrule
        XS & $140{,}456$ & 0.0562 & 0.0613 & 0.31 & 18 \\
        S & $300{,}612$ & 0.0632 & 0.0686 & 0.31 & 18 \\
        M & $849{,}732$ & 0.0684 & 0.0732 & 0.33 & 18 \\
        L & $963{,}587$ & 0.0949 & 0.0997 & 0.38 & 18 \\
        XL & $6{,}953{,}980$ & 0.592 & 0.600 & 2.89 & 78 \\
        XXL & $27{,}822{,}308$ & 1.674 & 1.686 & 5.20 & 78 \\
        \textsc{MACE-MP-0} (small) & $3{,}847{,}696$ & 15.5 & 26.6 & 23.7 & 78 \\
        \textsc{MACE-OMAT-0} (medium) & $9{,}063{,}204$ & 44.7 & 55.9 & 64.8 & 78 \\
        \bottomrule
    \end{tabular}
\end{table}

\begin{figure}[tbp]
    \centering
    \includegraphics[width=\linewidth]{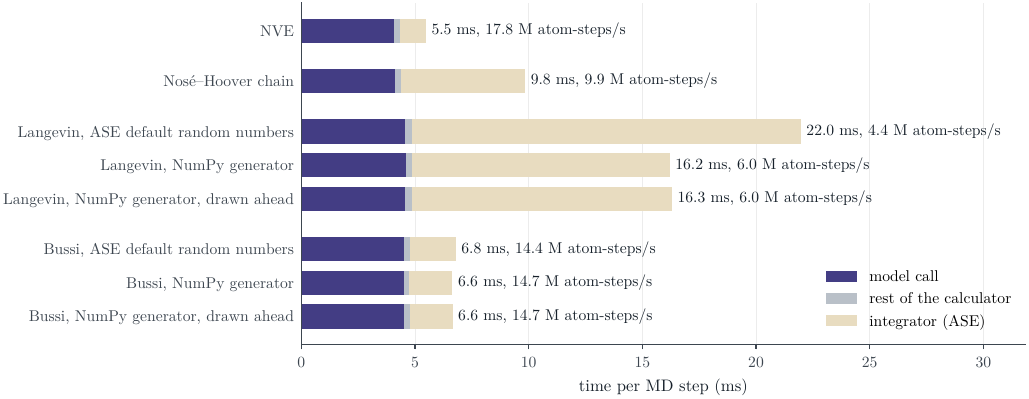}
    \caption{\textbf{Where the time of one MD step through \textsc{ASE} goes.}
    Per integrator and source of random numbers (\textsc{ILMlip} S, $97{,}556$~atoms of fcc Cu, one B200): the model call, the rest of the calculator (list check, results) and the integrator; at the end of each bar the step time and the throughput.
    The neighbour list is built in a background thread and overlaps the step.}
    \label{fig:si-md-steps}
\end{figure}

Outside the model call, the rest of the step splits into the calculator overhead (list check, waiting for a new list, results) and the integrator with its driver loop (Supplementary Figure~\ref{fig:si-md-steps}: velocity-Verlet, Nos\'e-Hoover chain, Langevin and Bussi).
With velocity-Verlet (NVE) the step of S on $97{,}556$~atoms takes $5.5$~ms ($17.8$~M atom-steps/s): the model call $4.1$, the calculator overhead $0.2$ and the integrator $1.2$~ms.
A thermostat shifts the balance.
With \textsc{ASE}'s Langevin thermostat the integrator takes $17.1$~ms with \textsc{ASE}'s default random numbers and $11.3$~ms with \texttt{default\_rng} ($11.5$~ms with the numbers of $100$~steps drawn at once), $3.7$ and $2.5$~times the model call, and the throughput falls to $4.4$ and $6.0$~M atom-steps/s.
The Bussi thermostat costs $1.9$ to $2.0$~ms in the integrator ($14.4$ to $14.7$~M atom-steps/s, $1.2$~times the NVE step) and a Nos\'e-Hoover chain $5.5$~ms ($9.9$~M).

\subsection*{Configurations of the calculator}
\label{si:speed-configs}

\begin{figure}[tbp]
    \centering
    \includegraphics[width=\linewidth]{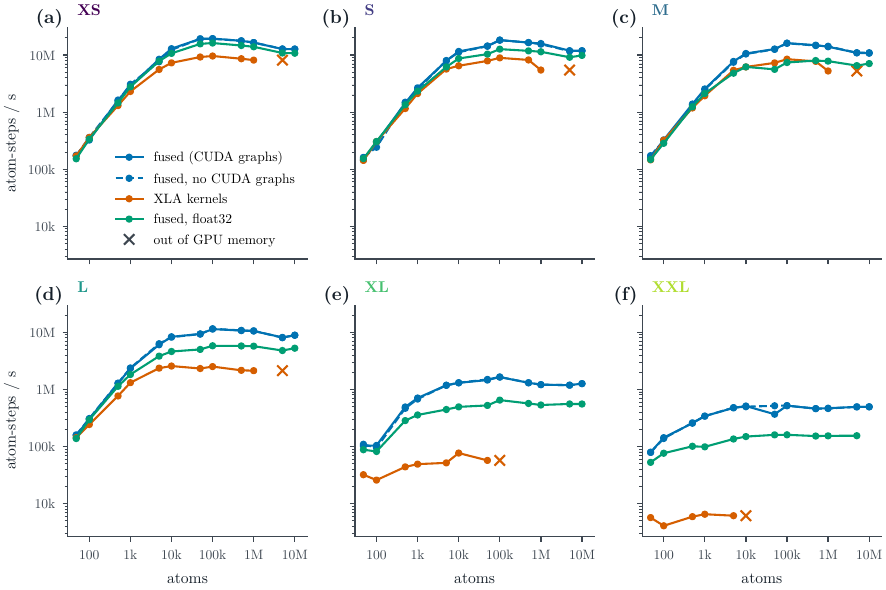}
    \caption{\textbf{MD throughput of each configuration of the \textsc{JAX} calculator per size.}
    Fused kernels with \textsc{CUDA} graphs, without graphs, XLA kernels, fused in float32; fcc Cu, one B200.
    A cross marks the first size that ran out of GPU memory.}
    \label{fig:si-speed-configs}
\end{figure}

Supplementary Figure~\ref{fig:si-speed-configs} compares the production calculator (fused kernels, \textsc{CUDA} graphs, TF32) with three alternatives on fcc Cu.
Without \textsc{CUDA} graphs the throughput stays within $4$~\% of production at $10^4$ and $10^5$ atoms in every size ($0.96$ to $1.02$).
Plain XLA operations instead of the fused kernels leave $0.56$ to $0.58$ of the throughput of XS to M at $10^4$ atoms, $0.31$ for L and $0.06$ for XL, and XL then no longer fits into memory at $10^5$ atoms and XXL at neither size.
Float32 instead of TF32 leaves $0.83$ for XS and $0.29$ for XXL at $10^4$ atoms: the larger the model, the larger its share of dense matrix products, which TF32 speeds up ($95$ and $97.5$~\% of the parameters of XL and XXL are in the readout, Supplementary Table~\ref{tab:si-sizes}).

\begin{figure}[tbp]
    \centering
    \includegraphics[width=\linewidth]{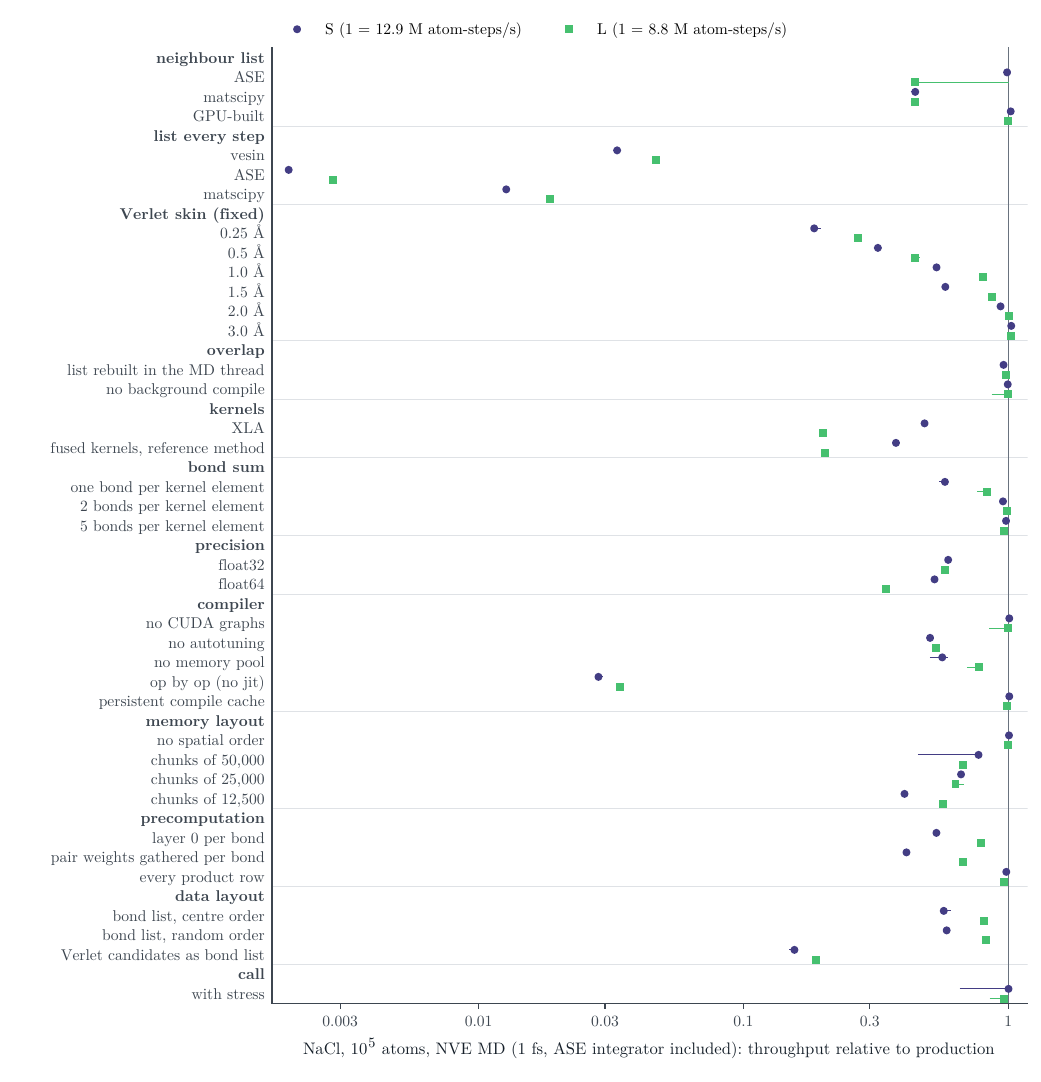}
    \caption{\textbf{Inference ablation.}
    MD throughput of each change to the production calculator relative to production, S and L, NaCl with $10^5$ atoms, NVE through \textsc{ASE}, one B200; bars: range over three rounds.}
    \label{fig:si-infer-abl}
\end{figure}

Starting from the production calculator, Supplementary Figure~\ref{fig:si-infer-abl} removes one of its features at a time, for S and L on NaCl with $10^5$ atoms in NVE through \textsc{ASE}, where production reaches $12.9$ and $8.8$~M atom-steps/s.
Rebuilding the neighbour list in every step leaves the following fractions of the production rate for S and L: $0.2$ and $0.3$~\% with the list of \textsc{ASE}, $1.3$ and $1.9$~\% with \textsc{matscipy}, $3.3$ and $4.7$~\% with \textsc{vesin}.
A skin of $0.25$ or $0.5$~{\AA} leaves $19$ to $45$~\%, because the list is then rebuilt after small displacements, while $3$~{\AA} does not differ from the production skin of $1.5$~{\AA}.
Running the model without compiling it leaves $2.8$ and $3.4$~\%, and a bond list instead of the candidate pairs $16$ and $19$~\%.
The kernels matter for both sizes but unequally: the reference form of the fused kernels leaves $38$~\% for S and $20$~\% for L, XLA operations $48$ and $20$~\% and double precision $53$ and $35$~\%.
Computing the stress, the space-filling-curve order, the compile cache, \textsc{CUDA} graphs and a list built on the GPU change nothing within the scatter.

\subsection*{CPU, laptop and \textsc{LAMMPS}}
\label{si:speed-cpu}

On CPUs \textsc{ILMlip} runs in a CPP engine in single precision (Supplementary Table~\ref{tab:si-cpu}; fcc Cu, velocity-Verlet through \textsc{ASE}).
On one thread of an \textsc{AMD EPYC} 9555 and $10^4$ atoms, XS and S reach $2.5$ and $1.6$~times the throughput of \textsc{NEP89} and L $0.65$~times; on $64$~threads of one process they reach $18$, $17$ and $6.4$~times, because \textsc{ILMlip} scales $25$ to $35$~times from one thread to $64$ and \textsc{NEP89} $3.4$~times.
The same S in \textsc{JAX} on the CPU reaches $32$~k on one thread, $0.46$~times the CPP engine.
The EAM potential runs on one thread only ($399$~k), so the comparison with S on $64$~threads ($6.1$~times) is one with a single-threaded reference.
On a laptop (\textsc{Intel} Core Ultra 5 125U) one core runs S at $1.7$~times and L at $0.68$~times the speed of \textsc{NEP89}.
\begin{table}[htbp]
    \caption{\textbf{CPU throughput for fcc Cu.}
    In $10^3$ atom-steps/s.
    EPYC: \textsc{AMD EPYC} 9555 ($64$ cores), one process with one or $64$~threads; laptop: \textsc{Intel} Core Ultra 5 125U, one process on one or two cores.
    EAM: \textsc{ASE} EAM calculator.
    \textsc{ILMlip} models: CPP engine, float32; S (\textsc{JAX}) is the XLA implementation on the CPU; --: not run.}
    \label{tab:si-cpu}
    \centering\small\setlength{\tabcolsep}{5pt}
    \begin{tabular}{lcccccccc}
        \toprule
        & \multicolumn{2}{c}{EPYC, 1 thread} & \multicolumn{2}{c}{EPYC, 64 threads} & \multicolumn{2}{c}{laptop, 1 core} & \multicolumn{1}{c}{laptop, 2 cores} \\
        & $10^3$ & $10^4$ & $10^3$ & $10^4$ & $10^4$ & $10^5$ & $10^4$ \\
        \midrule
        \textsc{ILMlip} XS & 111 & 106 & $2{,}240$ & $2{,}628$ & -- & -- & -- \\
        \textsc{ILMlip} S & 71.9 & 70.1 & $1{,}666$ & $2{,}422$ & 51.5 & 52.4 & 92.1 \\
        \textsc{ILMlip} L & 28.1 & 27.8 & 749 & 927 & 20.2 & 20.2 & 34.8 \\
        \textsc{ILMlip} S (\textsc{JAX}) & 36.3 & 32.0 & 125 & 172 & -- & -- & -- \\
        \textsc{NEP89} & 39.3 & 42.8 & 119 & 144 & 29.6 & 31.1 & 43.6 \\
        EAM & 406 & 399 & 407 & 399 & 225 & 187 & 222 \\
        \bottomrule
    \end{tabular}
\end{table}

In \textsc{LAMMPS}~\cite{lammps} \textsc{ILMlip} runs as an ML-IAP model on the KOKKOS GPU package~\cite{lammps_kokkos,kokkos} (Supplementary Table~\ref{tab:si-lammps}).
At $10^4$ atoms XS reaches $0.27$~times the throughput of the \textsc{ASE} calculator on the same card; at $10^6$ atoms it reaches $25.6$~M atom-steps/s, $1.5$~times its \textsc{ASE} value of $16.9$~M.
Lennard-Jones and EAM reach $942$ and $409$~M atom-steps/s at $10^6$ atoms, $37$ and $16$~times XS.
L needs three MPI ranks sharing the GPU to run $10^6$ atoms; XL and XXL ran out of memory with $7$ and $20$~ranks.
\begin{table}[htbp]
    \caption{\textbf{MD throughput in \textsc{LAMMPS}.}
    ML-IAP with KOKKOS, one B200, in $10^6$ atom-steps/s, and, for comparison, through \textsc{ASE} at $10^4$ atoms.
    Ranks: MPI ranks that share the GPU at $10^6$ atoms; OOM: out of memory.}
    \label{tab:si-lammps}
    \centering\small\setlength{\tabcolsep}{5pt}
    \begin{tabular}{lrrrr}
        \toprule
        & \textsc{ASE}, $10^4$ & \textsc{LAMMPS}, $10^4$ & \textsc{LAMMPS}, $10^6$ & ranks \\
        \midrule
        \textsc{ILMlip} XS & 12.5 & 3.38 & 25.6 & 1 \\
        \textsc{ILMlip} S & 11.6 & 3.17 & 21.9 & 1 \\
        \textsc{ILMlip} M & 10.6 & 3.08 & 18.7 & 1 \\
        \textsc{ILMlip} L & 8.02 & 2.91 & 10.5 & 3 \\
        \textsc{ILMlip} XL & 1.31 & 0.816 & OOM & 7 \\
        \textsc{ILMlip} XXL & 0.507 & 0.333 & OOM & 20 \\
        Lennard-Jones & -- & 119 & 942 & 1 \\
        EAM & -- & 22.2 & 409 & 1 \\
        \bottomrule
    \end{tabular}
\end{table}

\subsection*{Cost per hyperparameter}
\label{si:speed-hparam}

\begin{figure}[tbp]
    \centering
    \includegraphics[width=\linewidth]{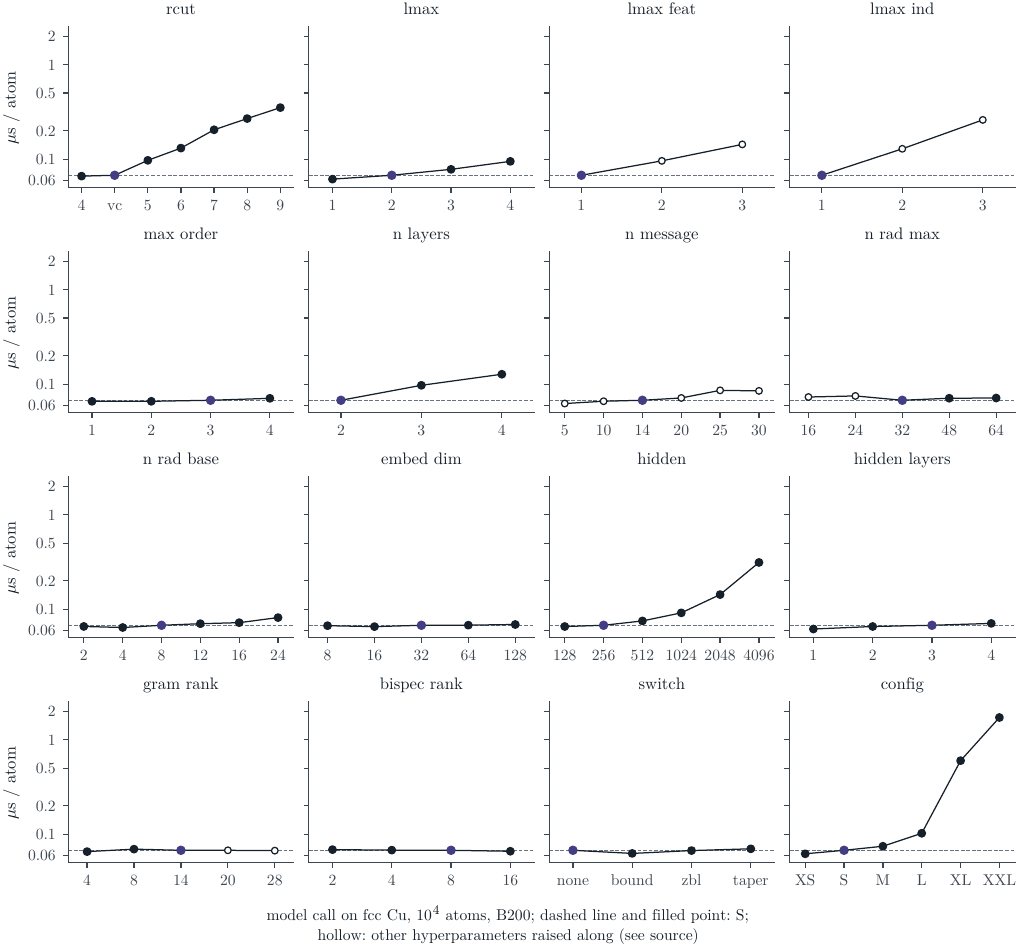}
    \caption{\textbf{Cost of one model call per hyperparameter.}
    Changed one at a time around S (fcc Cu, about $10^4$ atoms, random weights, one B200); dashed: S; open points: another hyperparameter had to change with it.
    The XL and XXL configurations use one $6$~{\AA} cut-off for all elements.}
    \label{fig:si-hparam-speed}
\end{figure}

The rule of the main text, to spend little per bond and freely per atom, rests on Supplementary Table~\ref{tab:si-hparam} and Supplementary Figure~\ref{fig:si-hparam-speed}: one model call on fcc Cu with about $10^4$ atoms and random weights, with one hyperparameter changed at a time around S.
The cut-off and the angular degrees of what travels along a bond dominate: a uniform cut-off of $6$~{\AA} instead of the element cut-offs of S raises the bonds per atom from $18$ to $78$, and $9$~{\AA} gives $248$.
Quantities per atom cost little, up to the readout width: $512$ channels (the readout of M) cost $1.11$~times, but $2{,}048$ and $4{,}096$ channels cost $2.1$ and $4.6$~times for $32$ and $120$~times the parameters of S.
The readouts of XL and XXL would add about $1.35$ and $2.1$~times to a model built like S; the rest of their cost per atom, $8.9$ and $25$~times S, comes from the cut-off, the degrees and the channel counts.
\begin{table}[htbp]
    \caption{\textbf{Cost of one model call relative to S when one hyperparameter changes.}
    S: $0.0676$~$\mu$s per atom; fcc Cu, $10^4$ atoms, random weights, one B200, stress on.
    The cut-off of S is element-dependent ($4$-$6$~{\AA}); the other cut-off values are one cut-off for all pairs.
    Where a value forces another to change (open points in Supplementary Figure~\ref{fig:si-hparam-speed}), the change is carried along: higher degrees of the message raise those of the features and of the pair object.
    $n_0$: channels of degree zero; $\nu$: number of factors of the product (body order $\nu+1$).
    Last rows: the released configurations, whose parameters and cost per atom include the hyperparameters of Supplementary Table~\ref{tab:si-sizes}.}
    \label{tab:si-hparam}
    \centering\small\setlength{\tabcolsep}{5pt}
    \begin{tabular}{llr@{\hspace{1.2em}}l}
        \toprule
        hyperparameter & S & \multicolumn{2}{l}{cost relative to S at the values tested} \\
        \midrule
        cut-off ({\AA}) & 4-6 & 5, 6, 7, 9 & 1.44, 1.94, 3.03, 5.20 \\
        $\ell_{\mathrm{msg}}$ & 1 & 2, 3 & 1.91, 3.85 \\
        readout width & 256 & 512, $1{,}024$, $2{,}048$, $4{,}096$ & 1.11, 1.35, 2.11, 4.62 \\
        $\ell_{\mathrm{feat}}$ & 1 & 2, 3 & 1.42, 2.12 \\
        layers & 2 & 3, 4 & 1.44, 1.89 \\
        $\ell_{\max}$ & 2 & 3, 4 & 1.16, 1.40 \\
        $n_{\mathrm{msg}}$ & 14 & 20, 25, 30 & 1.06, 1.27, 1.26 \\
        $n_{\mathrm{base}}$ & 8 & 12, 16, 24 & 1.04, 1.07, 1.21 \\
        $n_0$ & 32 & 48, 64 & 1.05, 1.06 \\
        $\nu$ & 3 & 4 & 1.05 \\
        readout depth & 3 & 4 & 1.05 \\
        element vector $d_z$ & 32 & 64, 128 & 1.00, 1.02 \\
        Gram rank $r_{\mathrm{G}}$ & 14 & 4, 8 & 0.97, 1.03 \\
        bispectrum rank $r_{\mathrm{B}}$ & 8 & 2, 4, 16 & 1.02, 1.00, 0.98 \\
        \midrule
        XS, S, M, L, XL, XXL & & & 0.92, 1.00, 1.10, 1.51, 8.88, 25.4 \\
        \bottomrule
    \end{tabular}
\end{table}

\subsection*{Many simulations in one call}
\label{si:speed-batched}

\begin{figure}[tbp]
    \centering
    \includegraphics[width=\linewidth]{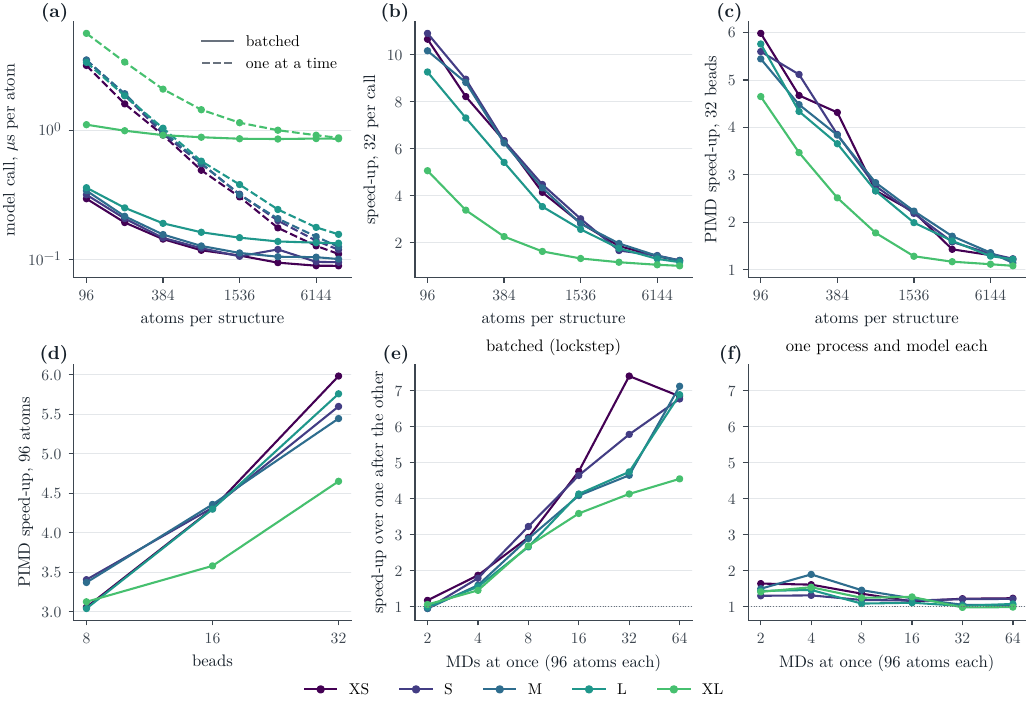}
    \caption{\textbf{Many structures per call.}
    Water, XS-XL, one B200: \textbf{(a)}~model-call time per atom, $32$~structures in one call against one at a time; \textbf{(b)}~the speed-up against structure size; \textbf{(c,~d)}~path-integral MD with \textsc{i-PI}, all beads in one call against one call per bead, against size and number of beads; \textbf{(e)}~$N$ MDs in lockstep with one batched call per step, and \textbf{(f)}~$N$ separate processes, as speed-up over running the $N$ MDs one after the other.}
    \label{fig:si-batched}
\end{figure}

Supplementary Table~\ref{tab:si-batched} and Supplementary Figure~\ref{fig:si-batched} measure the three ways of Methods to use one call for several structures, on water-like cells of XS to XL.
The gain is largest where one structure leaves the GPU idle: for S, $32$~structures of $96$~atoms in one call take $0.32$ instead of $3.5$~$\mu$s per atom, while at $9{,}216$~atoms per structure the gain falls to $1.02$-$1.25$, because one structure already fills the GPU adequately.
In path-integral MD with \textsc{i-PI}~\cite{ipi3} via an \textsc{ASE} socket, one call for all beads of a $96$-atom system is $3.4$, $4.3$ and $5.6$~times faster for S than one call per bead at $8$, $16$ and $32$~beads.
In lockstep MD, $64$ \textsc{ASE} simulations of $96$~atoms share one call per step: $4.0$~ms per step of all $64$ for S, against $22.5$~ms as $64$ separate processes on the same GPU.
\begin{table}[htbp]
    \caption{\textbf{Speed-up from evaluating many structures in one call.}
    Water-like cells, one B200.
    Call: $32$~structures of $96$ and of $9{,}216$~atoms in one call against one at a time.
    PIMD: one call for all beads against one call per bead, $96$~atoms, $8$ / $16$ / $32$~beads.
    $64$ MDs: $64$ \textsc{ASE} simulations of $96$~atoms, speed-up over running them one after another, in lockstep with one call per step and as separate processes ($64$ independent workers on the GPU).}
    \label{tab:si-batched}
    \centering\small\setlength{\tabcolsep}{5pt}
    \begin{tabular}{lrrlrr}
        \toprule
        & \multicolumn{2}{c}{call} & PIMD & \multicolumn{2}{c}{64 MDs} \\
        & 96 & $9{,}216$ & 8 / 16 / 32 beads & lockstep & processes \\
        \midrule
        XS & 10.7 & 1.23 & 3.06 / 4.30 / 5.98 & 6.8 & 1.23 \\
        S & 10.9 & 1.25 & 3.40 / 4.32 / 5.60 & 6.8 & 1.21 \\
        M & 10.2 & 1.24 & 3.37 / 4.36 / 5.45 & 7.1 & 1.05 \\
        L & 9.26 & 1.17 & 3.04 / 4.30 / 5.76 & 6.9 & 1.07 \\
        XL & 5.06 & 1.02 & 3.13 / 3.58 / 4.65 & 4.5 & 0.99 \\
        \bottomrule
    \end{tabular}
\end{table}

\subsection*{Training speed}
\label{si:speed-training}

Supplementary Table~\ref{tab:si-trainspeed} gives the throughput of the training step with the production recipe (Muon and AdamW, TF32, fused kernels, \textsc{CUDA} graphs).
The training set holds $15{,}625$~NaCl configurations of $64$~atoms (in total $10^6$~atoms).
XXL trains $5.0$~times more slowly than XS per atom, with $198$~times the parameters; the step compiles once, in $25$ to $53$~s.
In float32 S trains $13$~\% more slowly than in TF32, and with cells of $16$~atoms ($62{,}500$~configurations, in total $10^6$~atoms) instead of $64$~atoms $17$~\% faster.
\begin{table}[htbp]
    \caption{\textbf{Training throughput of the released model sizes on one B200.}
    NaCl cells of $64$~atoms, batches of $8{,}130$~atoms, production recipe, ten epochs of $123$~steps.}
    \label{tab:si-trainspeed}
    \centering\small\setlength{\tabcolsep}{5pt}
    \begin{tabular}{lrrrr}
        \toprule
        & atoms/s & seconds per epoch & peak memory (GiB) & compilation (s) \\
        \midrule
        XS & $648{,}793$ & 1.54 & 1.4 & 33.7 \\
        S & $617{,}024$ & 1.62 & 1.7 & 34.6 \\
        M & $602{,}946$ & 1.65 & 1.9 & 25.2 \\
        L & $486{,}574$ & 2.04 & 2.1 & 36.7 \\
        XL & $263{,}221$ & 3.75 & 9.5 & 53.2 \\
        XXL & $130{,}113$ & 7.51 & 18.6 & 48.7 \\
        \bottomrule
    \end{tabular}
\end{table}

\subsection*{Details of the GPU engine}

If a step finds more neighbours than the table holds, it is recomputed with a wider table, and the next width compiles in the background before it is needed, so that the table never silently drops a neighbour and rarely stalls the simulation.
Atoms are sorted along a space-filling curve~\cite{morton1966} to keep neighbours close in memory, and systems that would not fit into one call are split into chunks whose halo covers the reach of both layers, with results identical to one call.

A second detail concerns the fused kernels, which work through a batch of independent elements, by default one bond per element, and each element carries a fixed cost of about half a nanosecond on a B200 GPU, whatever arithmetic it holds, which dominates for a small model: the bond sum of \textsc{ILMlip}-S couples $14$ channels along $18$ paths per bond.
Because the dense neighbour table stores the bonds of one atom next to each other, we hand $G$ bonds of an atom to the kernel as one element, which repeats the polynomial $G$ times; every bond still gets the same products and the atom the same sum, while the number of elements falls $G$-fold.
With $G=4$ ($2$ for the largest model, whose $48$ channels and $118$ paths per bond already keep an element busy), the bond-sum kernel of \textsc{ILMlip}-S runs $3.3$~times faster forward and $2$~times faster with its backward pass, and the whole model call $13$-$20$~\% faster at $10^4$-$10^5$ atoms.

\clearpage
\section*{Supplementary Note 6: Accuracy of the released models}
\label{si:accuracy}

\subsection*{Errors on held-out data}
\label{si:acc-errors}

Errors are mean absolute errors (MAE), in meV/atom for the energy $E$, meV/{\AA} for the force components $F$ and meV/{\AA}$^3$ for the stress $S$.
Each source of OAM has a held-out part of $2$~\% split by shard: $208{,}128$~structures for the MPA models ($50{,}000$ \textsc{MPtrj}, $158{,}128$ \textsc{sAlex}) and $2{,}104{,}613$~structures from the $11$ subsets of \textsc{OMat24} for the OMAT models.
The \textsc{OMat24} hold-out is our own split, not the official validation set of \textsc{OMat24} ($1{,}074{,}643$~structures) on which \textsc{DPA-4.1C} and \textsc{NEP89} are scored in Ref.~\onlinecite{dpa-4c}, and the \textsc{NEP89} labels there carry a D3 correction while ours do not; the comparison rows of Supplementary Table~\ref{tab:si-err} give the order of magnitude, not a ranking.

\begin{table}[htbp]
    \caption{\textbf{Errors on held-out structures.}
    OAM-MPA on \textsc{MPtrj} and \textsc{sAlex}, OAM-OMAT on \textsc{OMat24} (own $2$~\% split).
    WiSE10: weight average of ten second-stage runs; committee5: mean prediction of five models.
    Lower rows: values of the \textsc{OMat24} validation set reported for other models in Ref.~\onlinecite{dpa-4c} (different set).}
    \label{tab:si-err}
    \centering\small\setlength{\tabcolsep}{5pt}
    \begin{tabular}{lrrrrrr}
        \toprule
        & \multicolumn{3}{c}{MPA} & \multicolumn{3}{c}{OMAT} \\
        & $E$ & $F$ & $S$ & $E$ & $F$ & $S$ \\
        \midrule
        \textsc{ILMlip}-OAM-XS & 28.66 & 73.9 & 4.153 & 26.86 & 123.2 & 5.240 \\
        \textsc{ILMlip}-OAM-S & 22.29 & 63.1 & 3.494 & 22.84 & 112.9 & 5.015 \\
        \textsc{ILMlip}-OAM-M & 19.47 & 59.2 & 3.317 & 20.78 & 108.1 & 4.773 \\
        \textsc{ILMlip}-OAM-L & 18.01 & 54.9 & 3.169 & 19.25 & 101.1 & 4.594 \\
        \textsc{ILMlip}-OAM-XL & 11.82 & 39.2 & 2.374 & 13.05 & 76.3 & 4.020 \\
        \textsc{ILMlip}-OAM-XXL & 10.65 & 40.1 & 2.551 & 11.62 & 67.3 & 3.916 \\
        \textsc{ILMlip}-OAM-L-WiSE10 & 17.97 & 54.9 & 3.125 & 19.14 & 101.1 & 4.636 \\
        \textsc{ILMlip}-OAM-XXL-WiSE10 & 10.59 & 38.6 & 2.421 & -- & -- & -- \\
        \textsc{ILMlip}-OAM-L-committee5 & 16.94 & 48.5 & 2.711 & 18.34 & 96.0 & 4.429 \\
        \midrule
        \textsc{NEP89} (D3 in labels) & & & & 85.5 & 269.9 & 6.7 \\
        \textsc{DPA-4.1C-Air} & & & & 19.3 & 105.4 & 3.6 \\
        \textsc{MACE-OMAT} medium & & & & 16.3 & 78.4 & 3.3 \\
        \bottomrule
    \end{tabular}
\end{table}

A weight average changes the errors of L by at most $1$~\% in $E$ and $F$ and by $-1.4$ to $+0.9$~\% in $S$; for XXL-MPA it lowers force and stress errors by $4$ and $5$~\%.
The committee lowers $E$, $F$, $S$ by $6$, $12$ and $14$~\% (MPA) and $5$, $5$ and $4$~\% (OMAT), at five calls.
Five calls of L cost $0.85$~times one call of XL, which has $30$ and $19$~\% lower energy and force errors than the committee (MPA).
The errors of \textsc{MatPES}, \textsc{OMol25} and \textsc{SolidAcid-0} models are in Supplementary Table~\ref{tab:si-training}.
The \textsc{OMol25} values are those of the validation split of the \textsc{OMol25} leaderboard ($2{,}762{,}021$~structures, no stress); its energy error per structure falls from $1{,}747$~meV for XS to $413$~meV for XL.
For XL it ranges from $203$~meV (neutral organics) to $487$~meV (biomolecules), and the force error from $15$ (biomolecules) to $56$~meV/{\AA} (metal complexes).
The \textsc{SolidAcid-0} errors are means over the $55$~acids of the per-acid MAE on $200$ test frames each, without stress; the energy errors of $0.62$ to $0.84$~meV/atom are lower and the force errors of $16.3$ to $20.5$~meV/{\AA} higher than those of the \textsc{MACE} medium model of the dataset paper~\cite{solidacid0}, $2.137$~meV/atom and $11.09$~meV/{\AA}.

\subsection*{Parity plots}
\label{si:acc-parity}

\begin{figure}[h]
    \centering
    \includegraphics[width=\linewidth]{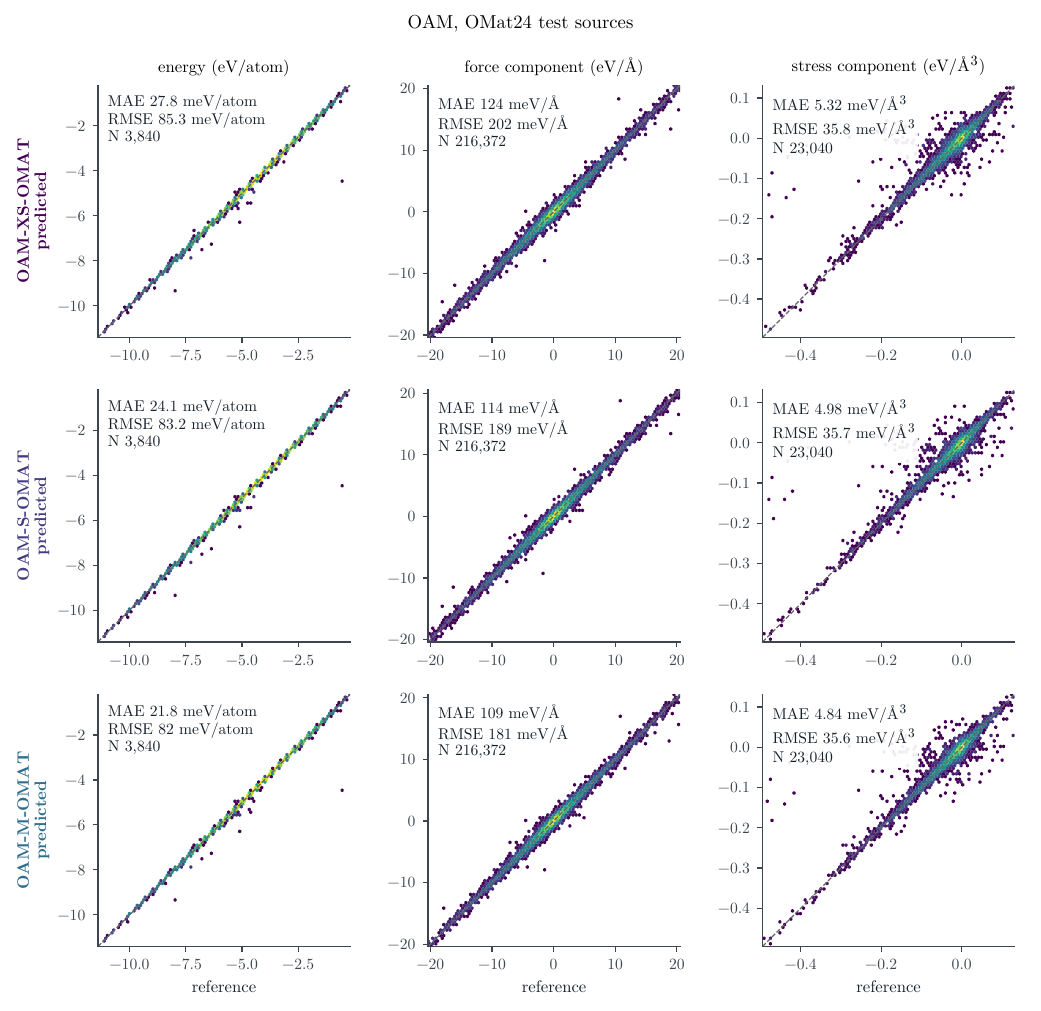}
    \caption{\textbf{Parity of energy, forces and stress on held-out structures.}
    OAM-OMAT models XS, S and M, with MAE and RMSE; continued in Supplementary Figure~\ref{fig:si-parity-oam_omat-2}.}
    \label{fig:si-parity-oam_omat}
\end{figure}

\begin{figure}[h]
    \centering
    \includegraphics[width=\linewidth]{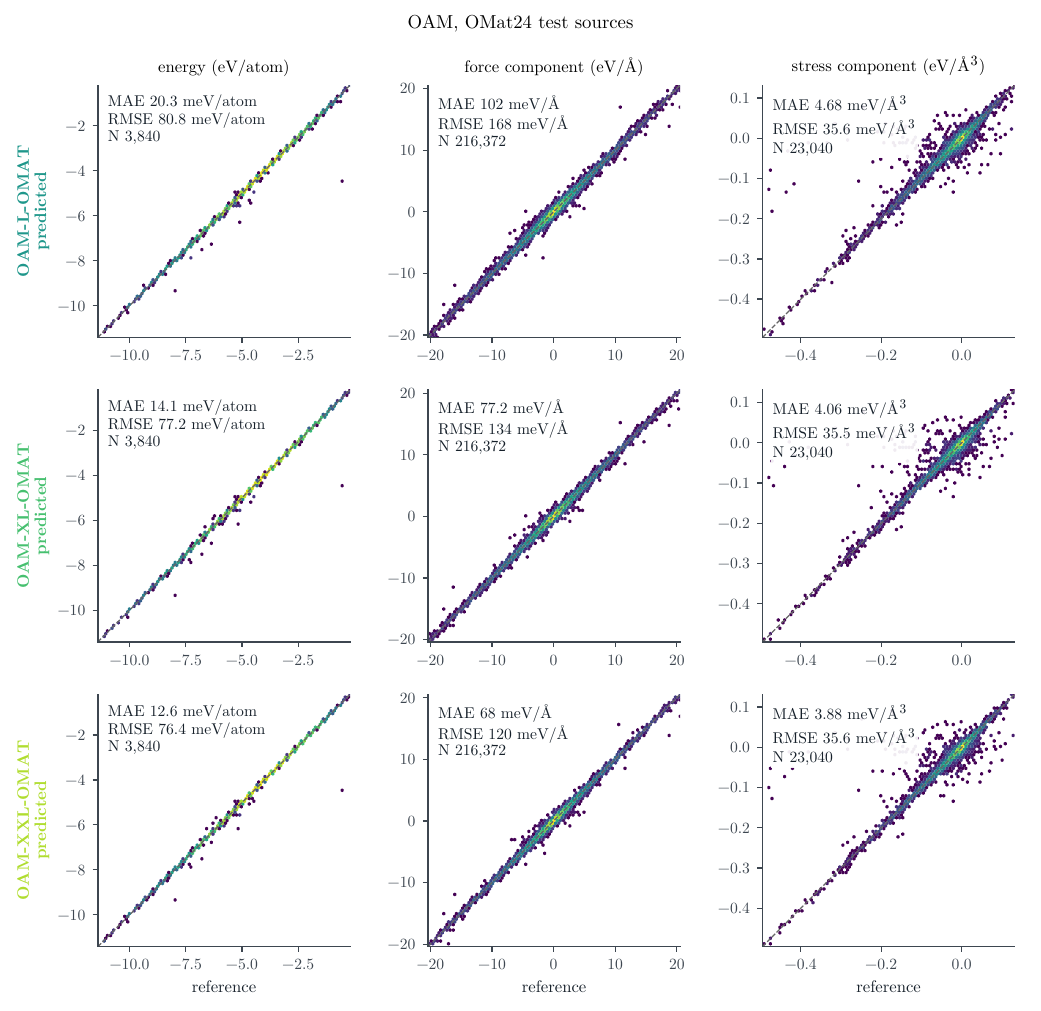}
    \caption{\textbf{Parity of energy, forces and stress on held-out structures (continued).}
    OAM-OMAT models L, XL and XXL, with MAE and RMSE; continued in Supplementary Figure~\ref{fig:si-parity-oam_omat-3}.}
    \label{fig:si-parity-oam_omat-2}
\end{figure}

\begin{figure}[h]
    \centering
    \includegraphics[width=\linewidth]{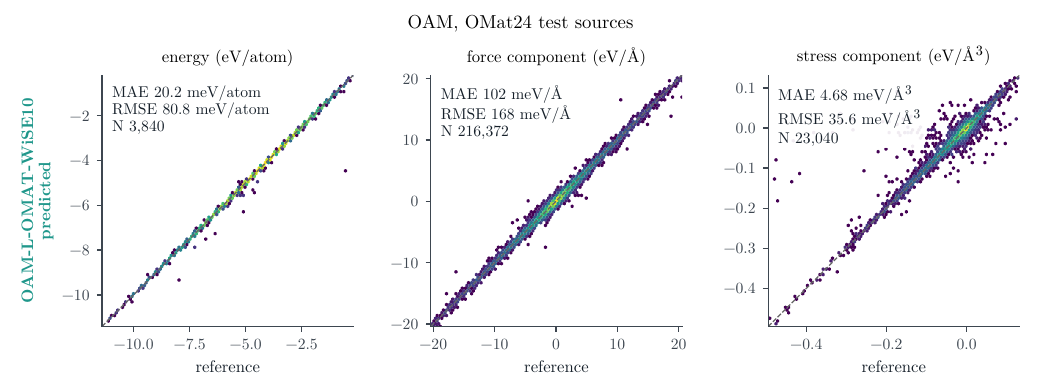}
    \caption{\textbf{Parity of energy, forces and stress on held-out structures (continued).}
    WiSE10 average of OAM-L-OMAT, with MAE and RMSE.}
    \label{fig:si-parity-oam_omat-3}
\end{figure}

\begin{figure}[h]
    \centering
    \includegraphics[width=\linewidth]{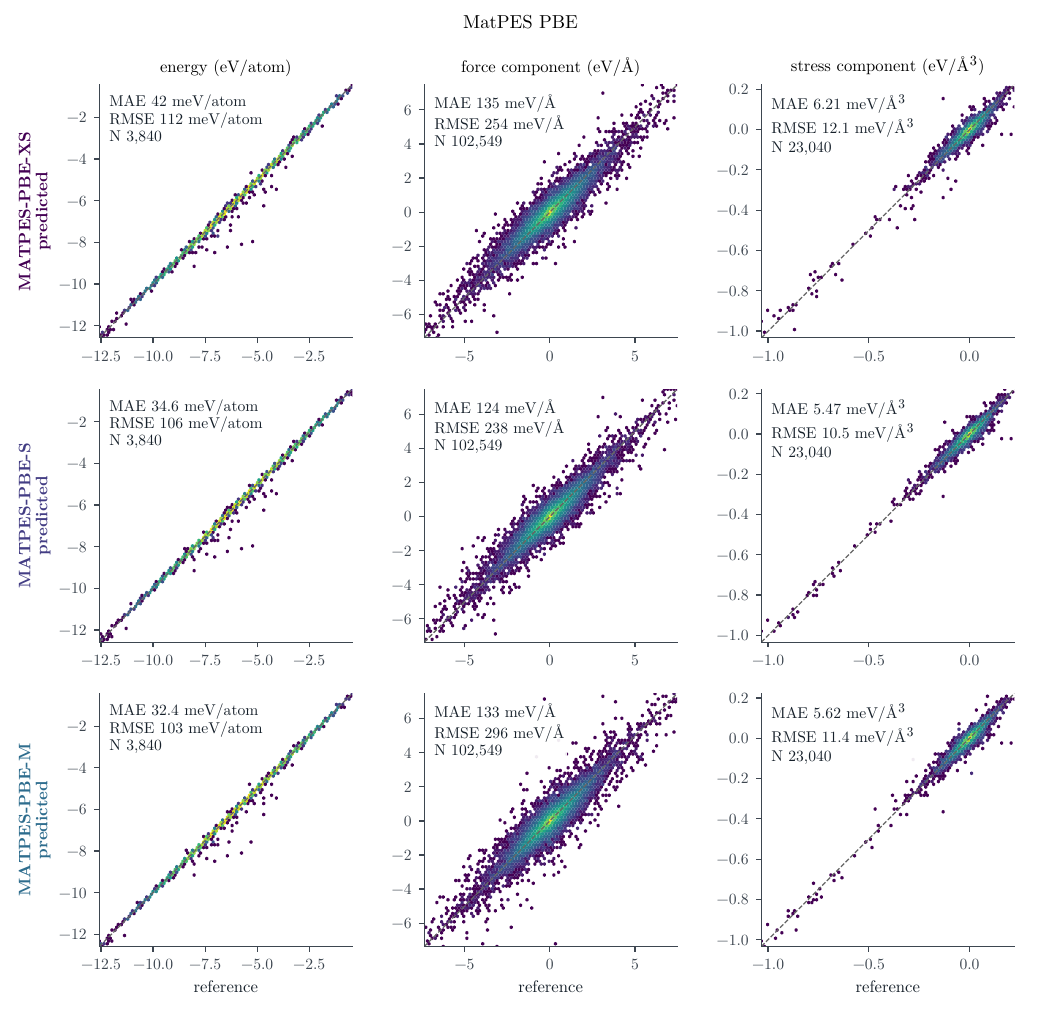}
    \caption{\textbf{Parity of energy, forces and stress on held-out structures.}
    \textsc{MatPES}-PBE models XS, S and M, with MAE and RMSE; continued in Supplementary Figure~\ref{fig:si-parity-matpes_pbe-2}.}
    \label{fig:si-parity-matpes_pbe}
\end{figure}

\begin{figure}[h]
    \centering
    \includegraphics[width=\linewidth]{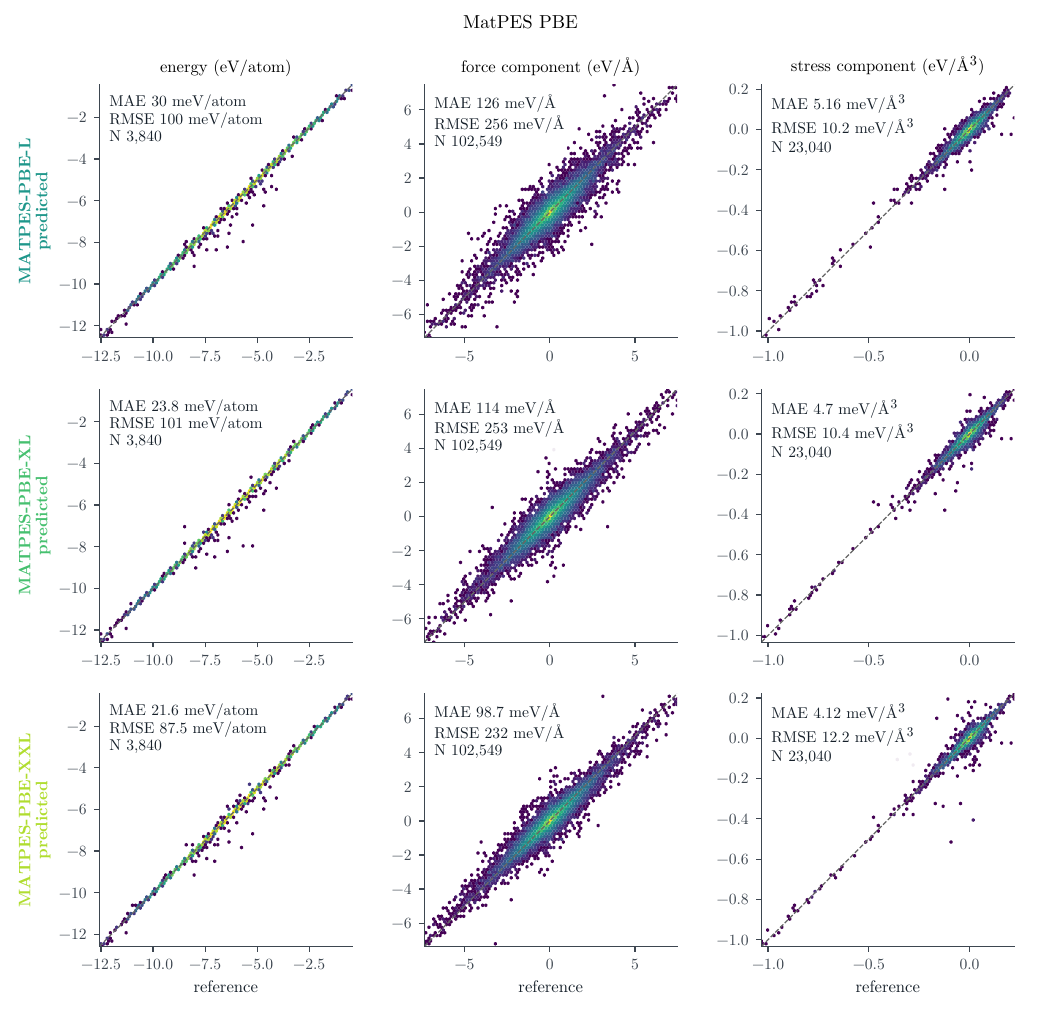}
    \caption{\textbf{Parity of energy, forces and stress on held-out structures (continued).}
    \textsc{MatPES}-PBE models L, XL and XXL, with MAE and RMSE.}
    \label{fig:si-parity-matpes_pbe-2}
\end{figure}

\begin{figure}[h]
    \centering
    \includegraphics[width=\linewidth]{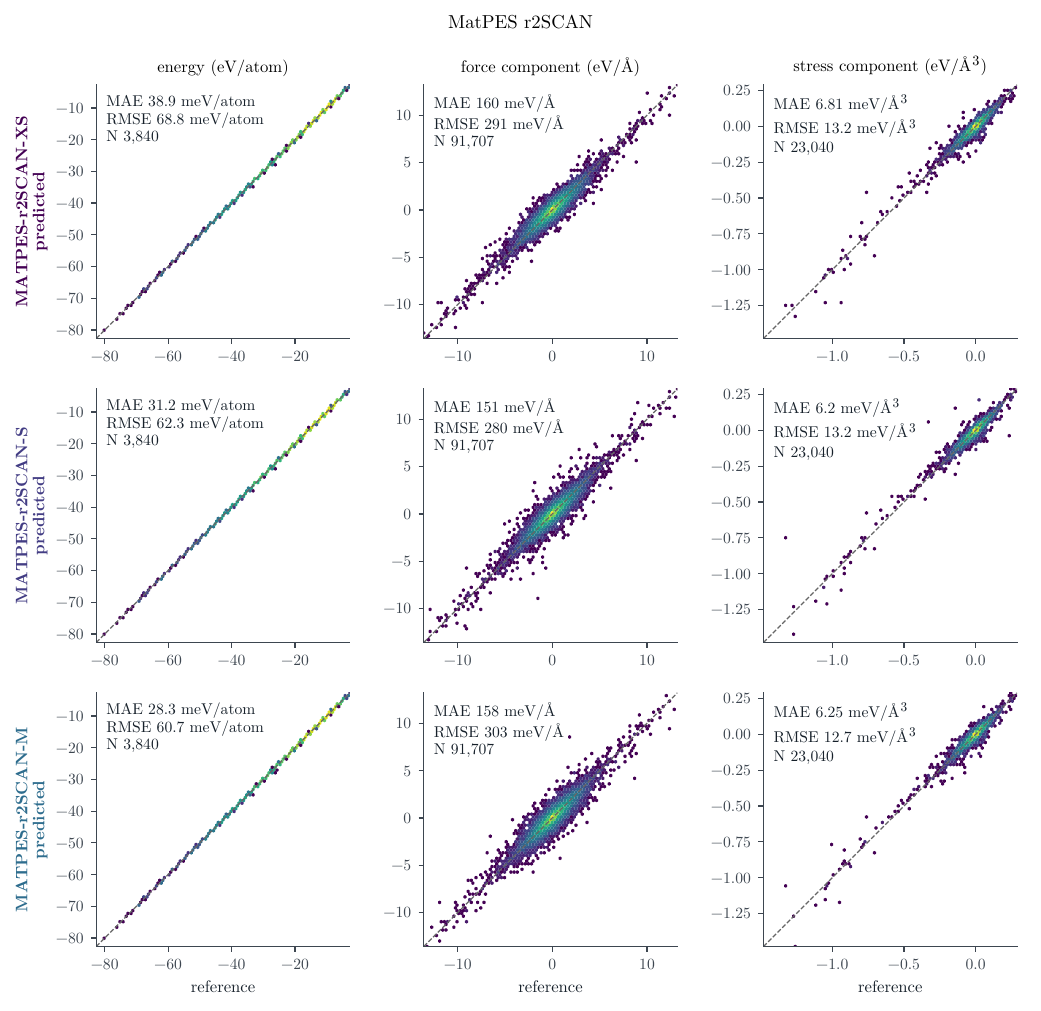}
    \caption{\textbf{Parity of energy, forces and stress on held-out structures.}
    \textsc{MatPES}-r$^2$SCAN models XS, S and M, with MAE and RMSE; continued in Supplementary Figure~\ref{fig:si-parity-matpes_r2scan-2}.}
    \label{fig:si-parity-matpes_r2scan}
\end{figure}

\begin{figure}[h]
    \centering
    \includegraphics[width=\linewidth]{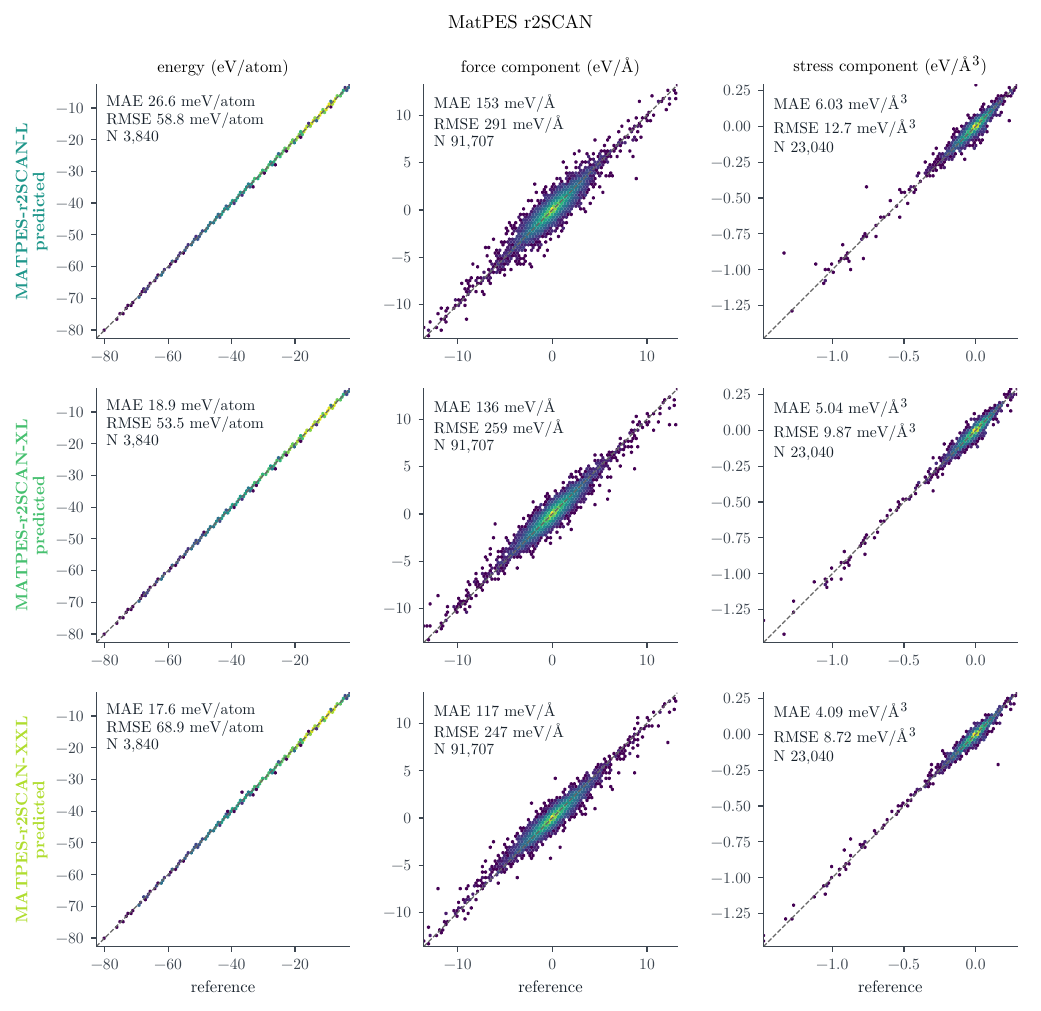}
    \caption{\textbf{Parity of energy, forces and stress on held-out structures (continued).}
    \textsc{MatPES}-r$^2$SCAN models L, XL and XXL, with MAE and RMSE.}
    \label{fig:si-parity-matpes_r2scan-2}
\end{figure}

\begin{figure}[h]
    \centering
    \includegraphics{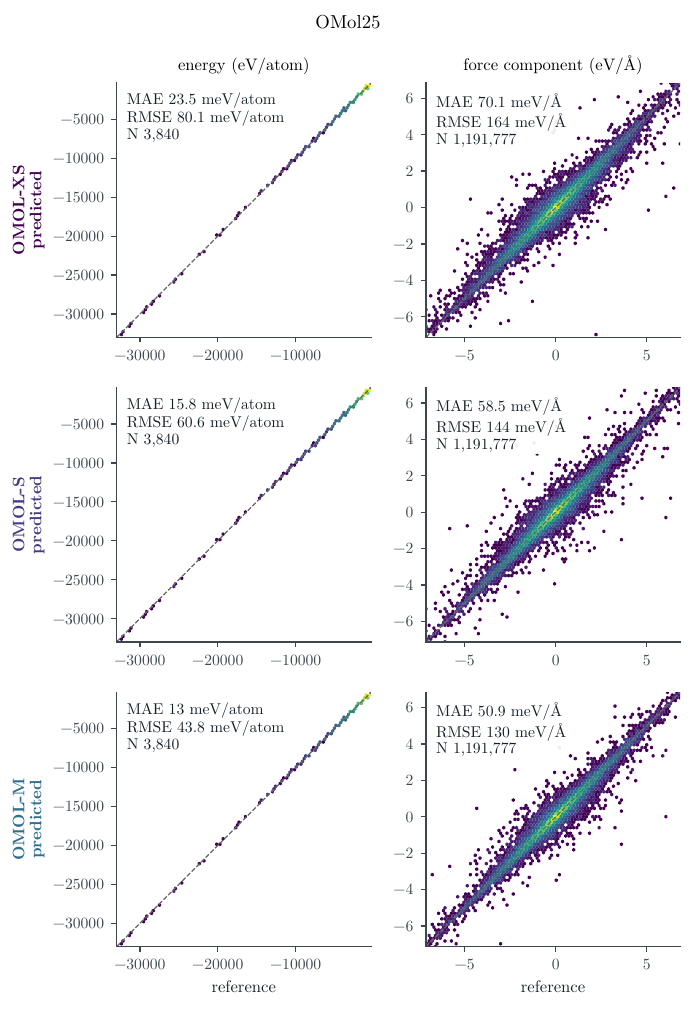}
    \caption{\textbf{Parity of energy and forces on held-out structures.}
    \textsc{OMol25} models XS, S and M, with MAE and RMSE; continued in Supplementary Figure~\ref{fig:si-parity-omol-2}.}
    \label{fig:si-parity-omol}
\end{figure}

\begin{figure}[h]
    \centering
    \includegraphics{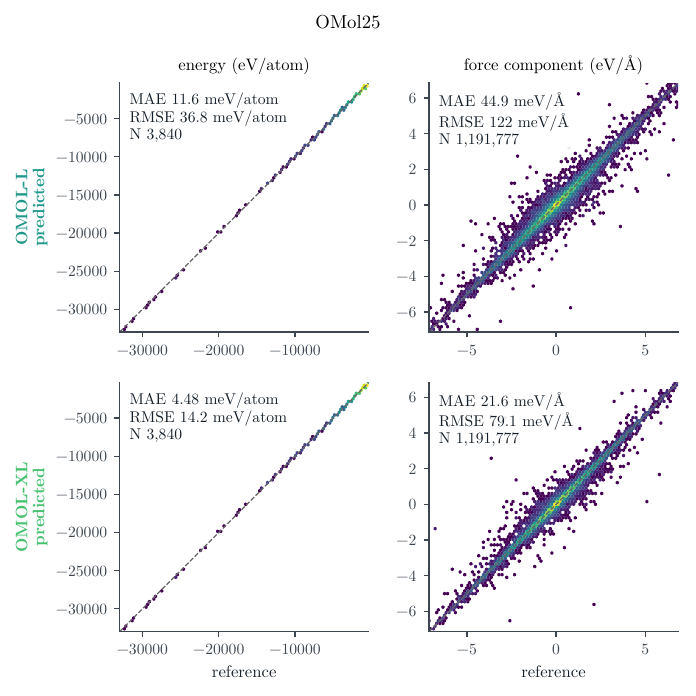}
    \caption{\textbf{Parity of energy and forces on held-out structures (continued).}
    \textsc{OMol25} models L and XL, with MAE and RMSE.}
    \label{fig:si-parity-omol-2}
\end{figure}

\begin{figure}[h]
    \centering
    \includegraphics{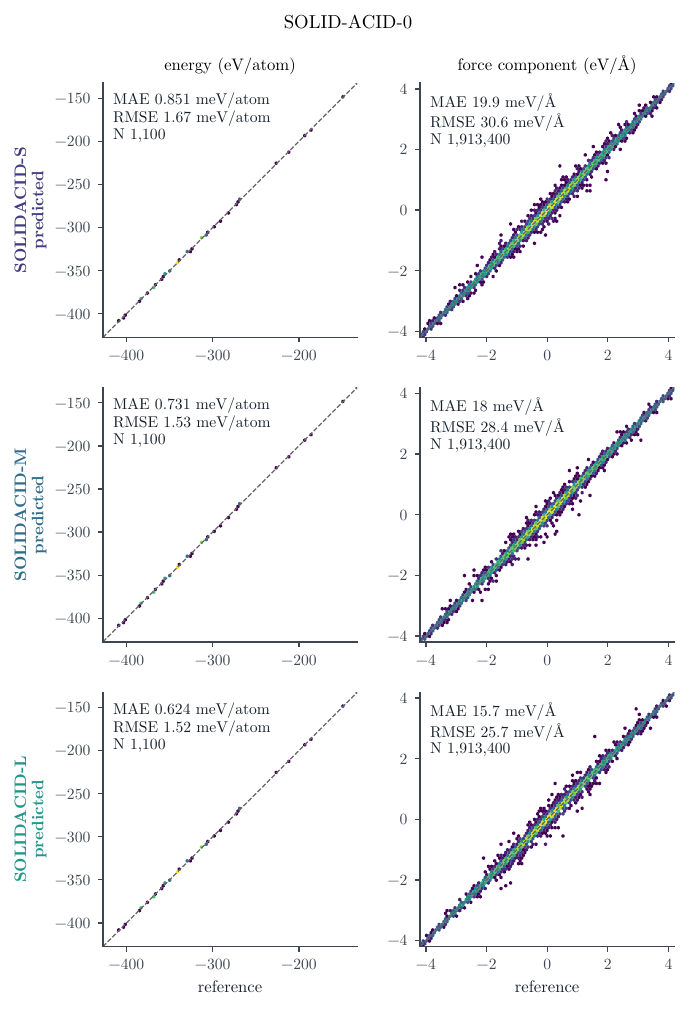}
    \caption{\textbf{Parity of energy and forces on held-out structures.}
    \textsc{SolidAcid-0} models, with MAE and RMSE.}
    \label{fig:si-parity-solidacid}
\end{figure}

Supplementary Figures~\ref{fig:si-parity-oam_omat}-\ref{fig:si-parity-solidacid} show energy, force and stress of every released model against the reference on a sample of $3{,}840$ held-out structures ($1{,}100$ for \textsc{SolidAcid-0}).
The OAM figures contain the OMAT models XS to XXL and the WiSE10 average of L; the MPA models and the committees are not shown.
The sample MAEs differ from the full test sets by up to $3$~meV/atom and $2$~meV/{\AA} for OAM-OMAT and \textsc{MatPES}-PBE, but are $7$ to $9$~meV/{\AA} lower in force for \textsc{MatPES}-r$^2$SCAN.

\clearpage
\subsection*{\textsc{Matbench} Discovery}
\label{si:acc-matbench}

Supplementary Table~\ref{tab:si-matbench} lists the metrics of the OAM-MPA models after relaxing the structures of WBM~\cite{wang2021wbm} with FIRE and a Fr\'echet cell filter to a maximum force of $0.02$~eV/{\AA} with the float32 CPP engine on CPUs.
F1~\cite{matbench} is that of the stability classification on the unique-prototype subset defined by the benchmark (the full set in a second column), $\kappa_{\mathrm{SRME}}$ is the symmetric relative mean error of the lattice thermal conductivity computed from the phonons of the model, and RMSD the root-mean-square displacement of the relaxed structures from the DFT ones in {\AA}.
\begin{table}[htbp]
    \caption{\textbf{\textsc{Matbench} Discovery metrics of the OAM-MPA models.}
    Epochs: first stage + second stage.
    WiSE10: ten second-stage runs averaged; committee5: five complete models.
    F1 on the unique prototypes and on the full test set (higher is better), RMSD in {\AA}, $\kappa_{\mathrm{SRME}}$ (lower is better).
    The submitted S is the $8+10$~epoch model.
    Rows below the rule: entries of the leaderboard repository (snapshot of November 2026).}
    \label{tab:si-matbench}
    \centering\small\setlength{\tabcolsep}{5pt}
    \begin{tabular}{lrcccccc}
        \toprule
        & parameters & epochs & F1 unique & F1 full & RMSD & $\kappa_{\mathrm{SRME}}$ \\
        \midrule
        \textsc{ILMlip}-OAM-XS-MPA & $140{,}456$ & 30 + 15 & 0.748 & 0.735 & 0.0831 & 0.752 \\
        \textsc{ILMlip}-OAM-S-MPA, 30 + 15 epochs & $300{,}612$ & 30 + 15 & 0.801 & 0.785 & 0.0780 & 0.533 \\
        \textsc{ILMlip}-OAM-S-MPA, 14 + 10 epochs & $300{,}612$ & 14 + 10 & 0.791 & 0.776 & 0.0782 & 0.588 \\
        \textsc{ILMlip}-OAM-S-MPA (submitted) & $300{,}612$ & 8 + 10 & 0.788 & 0.773 & 0.0793 & 0.630 \\
        \textsc{ILMlip}-OAM-M-MPA & $849{,}732$ & 30 + 15 & 0.827 & 0.809 & 0.0767 & 0.524 \\
        \textsc{ILMlip}-OAM-L-MPA & $963{,}587$ & 30 + 15 & 0.841 & 0.822 & 0.0734 & 0.503 \\
        \textsc{ILMlip}-OAM-L-MPA-WiSE10 & $963{,}587$ & 30 + 15 & 0.842 & 0.824 & 0.0735 & 0.507 \\
        \textsc{ILMlip}-OAM-L-MPA-committee5 & $4{,}817{,}935$ & 30 + 15 & 0.850 & 0.832 & 0.0710 & 0.419 \\
        \textsc{ILMlip}-OAM-XL-MPA & $6{,}953{,}980$ & 8 + 10 & 0.895 & 0.874 & 0.0642 & 0.281 \\
        \textsc{ILMlip}-OAM-XXL-MPA & $27{,}822{,}308$ & 8 + 10 & 0.911 & 0.893 & 0.0620 & 0.223 \\
        \textsc{ILMlip}-OAM-XXL-MPA-WiSE10 & $27{,}822{,}308$ & 8 + 10 & 0.910 & 0.892 & 0.0621 & 0.220 \\
        \midrule
        \textsc{DPA-4.1C-Air-OAM} & $433{,}673$ & & 0.799 & 0.783 & 0.0766 & 0.606 \\
        \textsc{MACE-MPA-0} & $9.06$~M & & 0.852 & 0.836 & 0.0731 & 0.412 \\
        \textsc{GRACE-2L-OAM} & $12.6$~M & & 0.880 & 0.861 & 0.0666 & 0.294 \\
        \textsc{DPA-4.1-Air-OAM} & $5.15$~M & & 0.921 & 0.902 & 0.0601 & 0.129 \\
        \textsc{DPA-4.1-Max-OAM-PFT} & $99$~M & & 0.935 & 0.914 & 0.0576 & 0.081 \\
        \bottomrule
    \end{tabular}
\end{table}

The sizes differ in training length, $30+15$~epochs for XS to L and $8+10$ for XL and XXL.
For S, F1 is $0.801$, $0.791$ and $0.788$ after $30+15$, $14+10$ and $8+10$~epochs, so XL and XXL are compared with a shorter-trained S; whether longer training raises their F1 we did not test.
The submitted S ($0.30$~million parameters, $6.9$~GPU-hours) trails \textsc{DPA-4.1C-Air-OAM} ($0.43$~million parameters, $19.8$~GPU-hours) in F1, $\kappa_{\mathrm{SRME}}$ and RMSD, whereas the S of $30+15$~epochs is above it in F1 and $\kappa_{\mathrm{SRME}}$ and below it in RMSD; M and L, with about twice its parameters, are $0.028$ and $0.042$ higher in F1.
The leader, \textsc{DPA-4.1-Max-OAM-PFT}, reaches $0.935$ with $99$~million parameters and $3{,}558$~GPU-hours; XXL reaches $0.911$ in $35.2$~GPU-hours, $1$~\% of that training time, and ranks $15$th of $74$ by unique-prototype F1 in our snapshot of the leaderboard~\cite{matbench} (November 2026).
The $\kappa_{\mathrm{SRME}}$ of XL is lower than that of \textsc{MACE-MPA-0} and \textsc{GRACE-2L-OAM} and higher than that of \textsc{DPA-4.1-Air-OAM}.
Averaging weights changes F1 by at most $0.001$ (L, XXL); the committee gains $0.009$ in F1 and $17$~\% in $\kappa_{\mathrm{SRME}}$ at the cost of five calls.

\clearpage
\subsection*{Reference-free certification}
\label{si:acc-pescert}

Supplementary Table~\ref{tab:si-pescert} lists the score of every \textsc{pescert} probe~\cite{haenseroth2026pescert} for the models of Fig.~2f of the main text, computed for ten elements (Li, Be, B, C, Na, Mg, Al, Si, P, S) with the CPP engine in double precision on CPUs.

\begin{table}[htbp]
    \caption{\textbf{Scores of the pescert probes.}
    Models of Fig.~2f of the main text ($1$ = ideal; the overall score is their geometric mean).}
    \label{tab:si-pescert}
    \centering\small\setlength{\tabcolsep}{5pt}
    \begin{tabular}{lrrrrrrrr}
        \toprule
        & XS & S & M & L & XL & XXL & L WiSE10 & L comm.~5 \\
        \midrule
        equivariance & 1.00 & 1.00 & 1.00 & 1.00 & 1.00 & 1.00 & 1.00 & 1.00 \\
        parity & 1.00 & 1.00 & 1.00 & 1.00 & 1.00 & 1.00 & 1.00 & 1.00 \\
        representation & 1.00 & 1.00 & 1.00 & 1.00 & 1.00 & 1.00 & 1.00 & 1.00 \\
        trimer & 0.96 & 0.69 & 0.67 & 0.77 & 0.76 & 0.55 & 0.68 & 0.76 \\
        zero modes & 0.59 & 0.55 & 0.52 & 0.47 & 0.56 & 0.74 & 0.46 & 0.45 \\
        Maxwell Betti & 1.00 & 1.00 & 1.00 & 1.00 & 1.00 & 1.00 & 1.00 & 1.00 \\
        conservativeness & 1.00 & 1.00 & 1.00 & 1.00 & 1.00 & 1.00 & 1.00 & 1.00 \\
        strain-position reciprocity & 0.91 & 0.98 & 0.97 & 0.86 & 0.92 & 0.89 & 0.86 & 0.96 \\
        reversibility & 1.00 & 1.00 & 1.00 & 1.00 & 1.00 & 1.00 & 1.00 & 1.00 \\
        stress consistency & 0.99 & 0.99 & 0.89 & 0.86 & 0.89 & 0.82 & 0.78 & 0.90 \\
        config.~temperature & 0.93 & 0.92 & 0.93 & 0.94 & 0.95 & 0.94 & 0.93 & 0.93 \\
        equipartition & 0.85 & 0.87 & 0.89 & 0.85 & 0.82 & 0.77 & 0.87 & 0.84 \\
        virial & 0.91 & 0.91 & 0.91 & 0.93 & 0.92 & 0.88 & 0.93 & 0.90 \\
        smoothness & 0.99 & 0.99 & 0.99 & 0.99 & 0.99 & 1.00 & 0.99 & 0.99 \\
        \midrule
        overall & 0.931 & 0.910 & 0.899 & 0.890 & 0.904 & 0.887 & 0.875 & 0.894 \\
        \bottomrule
    \end{tabular}
\end{table}

\clearpage
\section*{Supplementary Note 7: Fine-tuning}
\label{si:ft}

\subsection*{The six systems and the recipe}
\label{si:ft-setup}

The protocol of the Methods (Ref.~\onlinecite{atk}) uses the AIMD frames of Supplementary Table~\ref{tab:si-ft-systems}, a random $90/10$ split into training and validation, and errors on held-out trajectories that are not among these frames.
All four modes start from \textsc{ILMlip}-OAM-S-OMAT ($300{,}612$~parameters) with Adam at a peak learning rate of $5.2\times10^{-3}$ reached after a warm-up and decaying to about $1\times10^{-4}$, an exponential moving average of the weights with decay $0.99$, gradient clipping at $10$, float32, fused kernels and seed $0$; the reference energies are re-fitted and no stress labels are used.
LoRA has rank $8$ and trains $28{,}235$~parameters, naive and scratch $300{,}512$, multi-head $587{,}488$.
A replay set of $1{,}000$~structures labelled by the foundation model (weight $1$) is passed to the naive, LoRA and multi-head runs.
The $24$ runs trained for $4.24$~GPU-hours on B200 and H200 NVL nodes (naive $0.97$~h for the six systems, $183$ to $983$~s each; LoRA $0.83$, multi-head $1.49$, scratch $0.95$).
\begin{table}[htbp]
    \caption{\textbf{Data of the fine-tuning study.}
    Frames: AIMD frames in the data set; train / valid: random $90/10$ split; test: frames of the held-out trajectories.}
    \label{tab:si-ft-systems}
    \centering\small\setlength{\tabcolsep}{5pt}
    \begin{tabular}{llrrcr}
        \toprule
        system & material & frames & atoms per frame & train / valid & test frames \\
        \midrule
        CDP & CsH$_2$PO$_4$ & $2{,}000$ & $512$ & $1{,}800$ / 200 & 200 \\
        KOH & aqueous KOH & $2{,}000$ & $288$ & $1{,}800$ / 200 & 118 \\
        Li$_{13}$Si$_4$ & Li$_{13}$Si$_4$ alloy & $2{,}000$ & 204 & $1{,}800$ / 200 & 70 \\
        L-Pyro & L-pyroglutamic acid & $1{,}191$ & $144$ & $1{,}072$ / 119 & 200 \\
        MoS$_2$ & MoS$_2$ & $2{,}007$ & $106$ & $1{,}806$ / 201 & 200 \\
        PhOH & phenol in water & $2{,}000$ & $388$ & $1{,}800$ / 200 & 200 \\
        \bottomrule
    \end{tabular}
\end{table}

\subsection*{Errors of the four modes}
\label{si:ft-errors}

Supplementary Table~\ref{tab:si-ft-errors} lists the errors on the held-out trajectories of the four modes, of the foundation model before fine-tuning and of the five fine-tuned foundation models of Ref.~\onlinecite{atk}.
The force error is the mean norm of the force-vector error per atom, about twice the mean error per component that the fine-tuning code reports; the zero-shot errors use reference energies re-fitted by the same routine.
\begin{table}[htbp]
    \caption{\textbf{Errors of the fine-tuning modes on the held-out trajectories.}
    Force error (meV/{\AA}, mean norm of the force-vector error per atom; upper block) and energy error (meV/atom; lower block).
    Zero-shot: \textsc{ILMlip}-OAM-S-OMAT before fine-tuning; naive, LoRA, multi-head, scratch: the four modes from \textsc{ILMlip}-OAM-S-OMAT; \textsc{MACE} to \textsc{ORB}: the fine-tuned foundation models of Ref.~\onlinecite{atk}.
    Bold: lowest of the nine fine-tuned models in the row.
    Means over the six systems.}
    \label{tab:si-ft-errors}
    \centering\small\setlength{\tabcolsep}{5pt}
    \begin{tabular}{lcccccccccc}
        \toprule
        & zero-shot & naive & LoRA & multi-head & scratch & \textsc{MACE} & \textsc{GRACE} & \textsc{SevenNet} & \textsc{MatterSim} & \textsc{ORB} \\
        \midrule
        \multicolumn{11}{l}{force error (meV/{\AA})} \\
        CDP & 278.9 & 34.6 & 38.8 & 55.3 & 34.2 & 35.5 & 63.1 & \textbf{31.7} & 32.0 & 37.8 \\
        KOH & 152.1 & 40.4 & 44.3 & 59.2 & 41.5 & 35.1 & 48.5 & \textbf{19.3} & 35.4 & 42.6 \\
        Li$_{13}$Si$_4$ & 117.3 & 26.3 & 30.9 & 38.8 & 24.1 & 22.0 & 19.0 & \textbf{15.1} & 31.3 & 32.7 \\
        L-Pyro & 518.3 & 40.7 & 55.2 & 70.3 & 44.3 & 33.3 & 40.3 & \textbf{21.1} & 23.2 & 40.4 \\
        MoS$_2$ & 327.5 & 27.1 & 33.9 & 40.8 & 26.5 & 30.0 & 23.0 & 28.4 & \textbf{17.5} & 63.0 \\
        PhOH & 222.5 & 32.1 & 38.2 & 53.4 & 29.2 & \textbf{26.1} & 48.5 & 29.9 & 54.7 & 38.3 \\
        mean & 269.4 & 33.5 & 40.2 & 53.0 & 33.3 & 30.3 & 40.4 & 24.3 & 32.4 & 42.5 \\
        \midrule
        \multicolumn{11}{l}{energy error (meV/atom)} \\
        CDP & 0.93 & 0.25 & 0.20 & 0.32 & \textbf{0.16} & 0.41 & 0.17 & 0.96 & 0.51 & 1.92 \\
        KOH & 2.58 & 1.99 & 1.98 & \textbf{1.95} & 1.98 & 3.43 & 2.13 & 2.16 & 5.94 & 2.72 \\
        Li$_{13}$Si$_4$ & 1.17 & \textbf{0.12} & 0.24 & 0.22 & 0.13 & 3.08 & 1.43 & 1.86 & 2.34 & 4.05 \\
        L-Pyro & 1.63 & 0.18 & 0.20 & 0.27 & \textbf{0.12} & 1.52 & 0.13 & 1.83 & 3.20 & 0.75 \\
        MoS$_2$ & 1.90 & 0.17 & 0.15 & 0.46 & \textbf{0.12} & 1.09 & 0.43 & 0.13 & 0.24 & 3.66 \\
        PhOH & 0.72 & \textbf{0.12} & 0.13 & 0.18 & 0.19 & 0.53 & 0.22 & 0.14 & 4.62 & 1.50 \\
        mean & 1.49 & 0.472 & 0.483 & 0.567 & 0.450 & 1.677 & 0.752 & 1.180 & 2.808 & 2.433 \\
        \bottomrule
    \end{tabular}
\end{table}

\subsection*{Production molecular dynamics}
\label{si:ft-prod}

The tasks are those of the Methods: $9$ MD runs and a nudged elastic band of $22$ images for the S hop in MoS$_2$, each with the fine-tuned (naive) and the zero-shot S model.
Each MD run uses a Nos\'e-Hoover chain (\textsc{ASE}, damping time $50$~fs) with $0.5$~fs steps and a fixed centre of mass for up to $6{,}000{,}000$~steps ($3$~ns), a frame every $50$~fs, in float32 through the \textsc{JAX} calculator.
A $3$~ns run of the fine-tune takes $3.7$~h on one B200 ($445$~steps/s, $0.6$ to $2.3\times10^5$ atom-steps/s for cells of $144$ to $512$~atoms); all runs of the fine-tune used $41.7$~GPU-hours on a mix of GPUs (B200, H200, H100 NVL, A100, L40S).

The CDP diffusion coefficient depends on the fit window, so only values from the window of the AIMD reference are compared; with it the zero-shot model is already within a factor of $1.4$ of AIMD and the activation energies agree to $4$~\%.
In Li$_{13}$Si$_4$ fine-tuning moves $D$(Li) from three times too low to within $6$~\% of AIMD.
The NEB of the zero-shot model ends $0.43$~eV below its start (last image $-0.427$~eV).
One observable moves away from the reference with fine-tuning: in KOH the zero-shot model gives the water diffusion $8$~\% below AIMD and the fine-tune $55$~\% above it; the AIMD trajectory ends at $163.5$~ps, which limits the window.

\clearpage
\section*{Supplementary Note 8: Zero-shot checks}
\label{si:zero}

Unless a size is named, the model is \textsc{ILMlip}-OAM-L-OMAT; the \textsc{AM26} benchmark was also run with XL and XXL.

The melting points of seven materials are compared with experiment and with the job archive of \textsc{MeltBench}~\cite{meltbench}, downloaded on 21~November 2026 (protocol in the Methods): the temperature interval in which the interface of the two-phase cell switches from growth to melting brackets $T_{\mathrm m}$ (Supplementary Table~\ref{tab:si-melt}).
Only Al, Cu and NaCl are part of the \textsc{MeltBench-10} benchmark, the other four materials have two to seven archive entries.
The mean absolute error is $58.4$~K; six melting points lie $10$ to $113$~K below experiment and that of LiCl $64$~K above.
For Ni, Na and NaCl the experimental value lies at the upper edge of the bracket; the brackets are $18$ to $86$~K wide, and differences between potentials below this are not resolved.
On Al, Cu and NaCl, the error of $56.7$~K is lower than that of ten of the twelve other models with entries for all three and higher than that of \textsc{DPA-3.2-5M-OMat24} ($22.2$~K) and \textsc{SevenNet-Omni-i12} ($35.3$~K).
\begin{table}[htbp]
    \caption{\textbf{Melting points from solid-liquid coexistence with \textsc{ILMlip}-OAM-L-OMAT.}
    Bracket: temperature interval in which the phases switch from crystal growth to melting. 
    Archive: number of entries in the \textsc{MeltBench} job archive (downloaded 21~November 2026) and their median (minimum-maximum); the experimental values are those of the archive.}
    \label{tab:si-melt}
    \centering\small\setlength{\tabcolsep}{5pt}
    \begin{tabular}{lrrrcl}
        \toprule
        & experiment (K) & \textsc{ILMlip} (K) & error (K) & bracket (K) & archive entries \\
        \midrule
        Al & 933 & 873.5 & $-59.5$ & 839.7-886.4 & 45: 936 (559-$1{,}100$) \\
        Cu & $1358$ & 1267.5 & $-90.5$ & 1222.2-1290.1 & 28: $1{,}313$ ($1{,}151$-$1{,}425$) \\
        Ni & $1728$ & 1675.7 & $-52.3$ & 1641.6-1728.0 & 3: $1{,}725$ ($1{,}686$-$1{,}749$) \\
        Na & 371 & 361.5 & $-9.5$ & 352.5-371.0 & 2: 331 (311-351) \\
        NaCl & $1074$ & 1053.9 & $-20.1$ & 1020.3-1074.0 & 52: $1{,}062$ (794-$1{,}432$) \\
        KCl & $1043$ & 929.7 & $-113.3$ & 886.6-938.7 & 7: $1{,}009$ (922-$1{,}280$) \\
        LiCl & 883 & 946.6 & $+63.6$ & 927.2-971.3 & 5: 930 (812-$1{,}054$) \\
        \bottomrule
    \end{tabular}
\end{table}

Stability at high temperature is tested by the heating test of \textsc{MLIPArena}~\cite{mliparena}: $120$ random mixtures of elements are heated from $300$ to $3{,}000$~K, and the score is the area under the curve (AUC) of the fraction of simulations that complete.
\textsc{ILMlip}-OAM-L-OMAT completes $119$ of $120$ (AUC $0.977$), second of the $11$~models behind \textsc{ORBv2} ($0.985$); the published models range from $0.654$ (\textsc{M3GNet}) to $0.985$.

Amorphous structures are the subject of \textsc{AM26}~\cite{fragapane2026am26}, which scores a model by the relative RMSE (RelRMSE) on five amorphous materials: the root-mean-square error divided by the standard deviation of the reference values, in per cent.
Supplementary Table~\ref{tab:si-am26} gives the force RelRMSE per set; a-C is the hardest set at every size.
XXL ($18.6$~\%) lies between \textsc{MACE-MH-1} ($17.8$) and \textsc{MatRIS-10M-MP} ($20.6$), L ($31.2$) near \textsc{MACE-MP-0b3} ($33.6$); the best published value is $8.2$ (\textsc{UMA-M-1P1}).
The energy RelRMSE without removing a per-set offset, the convention of the benchmark, has a mean of $37.2$, $13.6$ and $11.1$~\% for L, XL and XXL.
\begin{table}[htbp]
    \caption{\textbf{Zero-shot force RelRMSE (\%) on the \textsc{AM26} sets.}
    The mean over the sets.}
    \label{tab:si-am26}
    \centering\small\setlength{\tabcolsep}{5pt}
    \begin{tabular}{lrrrrrrr}
        \toprule
        & a-C & a-Si & a-SiO$_2$ & a-GST & a-LiPS & mean \\
        \midrule
        \textsc{ILMlip}-OAM-L-OMAT & 54.9 & 28.6 & 8.9 & 32.1 & 31.2 & 31.2 \\
        \textsc{ILMlip}-OAM-XL-OMAT & 37.5 & 18.7 & 4.6 & 23.5 & 22.1 & 21.3 \\
        \textsc{ILMlip}-OAM-XXL-OMAT & 33.9 & 15.3 & 3.9 & 20.4 & 19.2 & 18.6 \\
        \bottomrule
    \end{tabular}
\end{table}

\clearpage
\bibliography{references}